\documentclass[11pt]{article}
\newcommand{\HRule}{\rule{\linewidth}{0.5mm}}

\usepackage{url}
\usepackage{amsmath}
\usepackage{amsthm}
\usepackage{array}
\usepackage{eucal}
\usepackage{amssymb}
\usepackage{mathrsfs}
\usepackage{color}
\usepackage{graphicx}
\begin{document}

%\begin{titlepage}
\begin{center}

% Upper part of the page. The '~' is needed because \\
% only works if a paragraph has started.
%\includegraphics[width=0.15\textwidth]{../uc-davis}~\\[0.5cm]

%\textsc{\LARGE University of California, Davis}\\[1cm]

% Title
\HRule \\[0.4cm]
{ \huge \bfseries The Raspberry model for protein-like particles: ellipsoids and confinement in cylindrical pores \\[0.4cm] }

\HRule \\[1cm]

% Author and supervisor
\begin{minipage}{0.8\textwidth}
	\begin{flushleft} \large
		\emph{Authors:}\\
		Vincent D. \textsc{Ustach}, Roland \textsc{Faller}\\
		Department of Chemical Engineering, University of California Davis, One Shields Ave, Davis, CA 95616, USA
	\end{flushleft}
\end{minipage}

\begin{minipage}{0.4\textwidth}
\large \date{July 11, 2016}
\vfill
\end{minipage}

% Bottom of the page
\end{center}
%\end{titlepage}

\abstract{
The study of protein mass transport \textit{via} atomistic simulation requires\
time and length scales beyond the computational capabilities of modern\
computer systems. The raspberry model for colloidal particles\
in combination with the mesoscopic hydrodynamic method of lattice\
Boltzmann facilitates coarse-grained simulations that are on the order\
of microseconds and hundreds of nanometers for the study\
of diffusive transport of protein-like colloid particles. The raspberry\
model reproduces linearity in resistance to motion \textit{versus} particle\
size and correct enhanced drag within cylindrical pores at off-center\
coordinates for spherical particles. Owing to the high aspect ratio\
of many proteins, ellipsoidal raspberry colloid particles were\
constructed and reproduced the geometric resistance factors of Perrin and of Happel and Brenner\
in the laboratory-frame and in the moving body-frame.\
Accurate body-frame rotations during diffusive motion\
have been captured for the first time using projections of displacements.\
The spatial discretization of the fluid leads to a renormalization of the\
hydrodynamic radius, however, the data describes a\
self-consistent hydrodynamic frame within this renormalized system.\
} %end of abstract
%
%\maketitle
%

\section{Introduction}
\label{sec:intro}

The development of devices employing protein transport through complex\
geometries is a critical problem with a great deal left to be\
comprehended. An accurate simulation model of proteins that allows\
efficient simulations of large systems ($>50$~nm) would provide a\
platform for computational studies to obtain a deeper understanding\
of methods for protein detection~\cite{davenport2012,dekker2013},\
analysis~\cite{fologea2007}, drug delivery~\cite{lesinski2005}, and\
molecular separation~\cite{safty2011}, as well as processes on long\
time scales, such as protein oligomerization and the effects of\
polymer-protein conjugation on transport.

The motion of colloid particles is governed by surface\
interactions, volume exclusions, hydrodynamic interactions, and\
electrostatics~\cite{russel1989}. In confined spaces, such as within nanoporous\
media, there is a large surface area to volume ratio, and all\
aforementioned interactions become enhanced due to close proximities\
of immersed particles and boundaries. These enhanced interactions lead to\
complexity in the transport properties of colloids such as\
proteins. For complex geometries the mathematical description of a\
system becomes a major hurdle to understanding. A modern review of\
approximations has been published by Dechadilok and Deen~\cite{deen2006}.

Modeling is an effective way to understand the influence of\
environmental factors on transport, and here we choose molecular\
dynamics (MD) as our method. Studies in MD on diffusion of proteins\
is limited to mainly studies of cytoplasm\
~\cite{frembgen-kesner2009,ridgway2008,ando2010,dlugosz2011,mereghetti2012}\
or lipid membranes~\cite{javanainen2013,goose2013}. A limited number\
of studies simulating protein diffusion for separation and detection\
exist. In order to capture events that require long time scales with\
all-atom MD, such as nanopore translocation, steered trajectories are\
often required~\cite{kannam2014}. An attempt to capture the rotational\
diffusion tensor through autocorrelation measurements proved to require\
trajectories outside the timescales available to all-atom methods\
~\cite{wong2008}. Liang et al.~\cite{liang2012} simulated\
coarse-grained MARTINI models~\cite{monticelli2008} of human serum\
albumin and bovine hemoglobin proteins sorbing to ion-exchange\
chromatographic media, obtaining 4.80~$\mu$s of simulated time.\
Zavadlav et al.~\cite{zavadlav2014} demonstrated a protein solvated\
in water described by adaptive resolution, in which water was modeled\
as individual molecules close to the protein and as four molecules per\
interaction site at longer distances. More extreme Coarse-graining\
models include single bead models of proteins~\cite{lee2012,tringe2015chemphys},\
$\alpha-$helix models ~\cite{javidpour2009,yang2010}, or the use of\
combinations of continuum methods and explicit MD particles~\cite{jubery2012}.

This study seeks a model with a consistent set of input\
parameters that accurately reproduces the diffusive transport of\
proteins (sans hydrophobic and electrostatic interactions).\
An accurate and efficient\
non-steered MD model of proteins that\
includes all major interactions is the eventual goal.\
While atomic resolution is a desirable\
goal in the modeling of soft materials in complex geometries,\
accessible time and length scales limit the efficient implementation of\
a sufficiently detailed model. Therefore, coarse-graining is required\
to obtain quantitative descriptions.
We begin with a spherical model and subsequently augment the validation\
using ellipsoidal geometries. Modeling proteins as ellipsoids is a common\
tactic in experimental~\cite{roosen-runge2011,yaroslav2006}\
and modeling~\cite{kovalenko2006,schluttig2010} contexts.

The intended use of the presented coarse-grained model is to increase\
the understanding in mass-transport processes in proteins. As such,\
measurement of diffusive transport is the method by which we validate\
the model. We discuss diffusion in the context of translation and\
rotation. The average properties of the Brownian motion of a particle\
is described by the diffusivity, the relationship between the driving\
force of thermal energy and the frictional resistance to motion~\cite{einstein1905}.\
Diffusion is often measured experimentally using, for example,\
particle tracking~\cite{han2006}, fluorescence correlation\
spectroscopy~\cite{magde1972}, scattering~\cite{roosen-runge2011}, or NMR\
relaxation~\cite{yaroslav2006}.

Without knowledge of the orientational\
configuration of the body, diffusive motion appears isotropic in the\
static laboratory coordinate frame (lab-frame). The lab-frame\
diffusivity is characterized by a scalar, $D$. Diffusivity can be more\
completely be described as a tensor $\overline{\overline{D}}$. The\
diffusivity tensor is a symmetric 3-by-3 matrix, and in the coordinate\
frame defined by the principal axes of the particle body, also known\
as the body-frame, the diffusivity tensor becomes diagonal:

\begin{equation}
	\label{eq:diffusion_tensor}
	\overline{\overline{D}} = \overline{\overline{\mu}}k_{B}T
\end{equation}

\begin{equation}
	\label{eq:diffusion_tensor_diagonal}
	\begin{pmatrix}
	D_{x^{b}} & 0 & 0 \\[0.3em]
	0 & D_{y^{b}} & 0 \\[0.3em]
	0 & 0 & D_{z^{b}} \\[0.3em]
	\end{pmatrix} =
	\begin{pmatrix}
	1/\xi_{col,x^{b}} & 0 & 0 \\[0.3em]
	0 & 1/\xi_{col,y^{b}} & 0 \\[0.3em]
	0 & 0 & 1/\xi_{col,z^{b}} \\[0.3em]
	\end{pmatrix} k_{B}T
\end{equation}

\noindent
where $\overline{\overline{\mu}}$ is the mobility tensor and\
$\xi_{col,x^{b}}$, $\xi_{col,y^{b}}$, $\xi_{col,z^{b}}$ are the resistance\
values in the body-frame that correspond to the principal diffusivity\
values $D_{x^{b}}$, $D_{y^{b}}$, $D_{z^{b}}$. For spherical particles,\
all diagonal elements are equal and the tensor collapses to a scalar value.

Many processes on the mass-transport timescale of proteins rely on\
anisotropic diffusion. The timescale for crossover from anisotropic\
to isotropic diffusion depend on the rotational dynamics of the\
particle~\cite{han2006}. For proteins, this timescale is on the order\
of microseconds, therefore, the translation of a protein through a\
pore may be dominated by anisotropic diffusion or, in the case of small\
pores, require a body-frame rotation to proceed. Information on anisotropic rotation\
is needed to understand NMR relaxation~\cite{yaroslav2006},\
fluorescence spectroscopy, protein oligomerization~\cite{kuttner2005}.\
The rotational reorientation needed to enter a small pore may help\
explain the strange behavior of pore diffusivity of proteins~\cite{ku2004}. Ensuring a\
complete description of the correct diffusion tensor for particles in our model is\
therefore critical.

We validate the anisotropic motion based on the\
calculations of Perrin~\cite{perrin1934,perrin1936} and Happell \&\
Brenner~\cite{happelbrenner1983}. In those works, the authors\
investigated the flow field past an ellipsoid and determined the\
disturbance to flow has the same form as a sphere with a radius\
$R_{eff}$. The relationship between this effective radius $R_{eff}$ and\
the dimensions of the ellipsoid result are the geometric factors that\
we validate our results against in this study.

The motion of a colloid particle in a pore of comparable size is\
affected by the hydrodynamic coupling enforced by no-slip boundary conditions between the particle\
and the wall~\cite{deen2006}. The result is that the resistance to motion will be\
higher. The increased resistance is called enhanced drag and (for a sphere) depends on\
the direction of motion, the ratio of particle size to pore size,\
and the distance to the pore wall. Many protein separation processes involve\
concentration gradient-driven transport through pores. We therefore\
seek to validate the diffusive transport of colloid particles through\
cylindrical pores and reproduce the enhanced drag calculated previously~\cite{higdon1995}.

The present study simulates protein-like colloid particles using the\
raspberry model scheme~\cite{dunweg2004} in combination with a fast\
hydrodynamic solver, the lattice Boltzmann (LB) method~\cite{succi,dunweg2008}, to\
explore the model at the protein scale and to study anisotropic\
transport due to particle aspect ratio and confinement. This contribution\
also sets the foundation for future model development that\
will include electrostatic as well as hydrophobic interactions.

\subsection{Hydrodynamic Interactions}
\label{sec:hydrodynamics}
The straightforward way to model hydrodynamic interactions is\
through the explicit treatment of water. In large systems, most of the\
computational effort is spent on water. Many methods have been devised to\
describe the effect of fluid interactions on particle motion\
without paying the cost of an explicit solvent, such as dissipative\
particle dynamics~\cite{dunweg2003} and Stokesian dynamics~\cite{brady1988}. In order to successfully model hydrodynamic interactions,\
the algorithm must exhibit Galilean invariance and recovery\
of the Navier-Stokes equations~\cite{succi}.

Originally developed as a lattice-gas solver~\cite{mcnamara1988}, the lattice Boltzmann method is a fluid phase model\
constructed from a discretization of the Boltzmann transport equation\
in space and time. LB reproduces the incompressible Navier-Stokes equation for mass\
and momentum transport at long length and timescales. The relaxation\
of fluid degrees of freedom in liquid systems is much faster than the\
transport of particles, and this separation of timescales allows\
the microscopic details of the fluid to be neglected~\cite{succi}. Compared to explicit\
solvents, LB is computationally efficient.

In LB the fluid ``lives" as density packets on a three dimensional\
square grid of nodes \textit{aka} the lattice. The fluid follows a\
two step scheme: the streaming step followed by the collision step.\
In the streaming step, fluid moves to neighboring nodes along\
discrete velocity links. In the collision step,\
inbound fluid densities at every node exchange momentum and relax\
towards an equilibrium that represents the Maxwell-Boltzmann velocity\
distribution. The simplest way to model relaxation is the lattice\
Bhatnagar, Gross, and Krook~\cite{LBGK1954} method, where the\
collision operator is simply a $1/\tau$ term, and the fluid relaxes over a single time scale.\
In the modern multi-relaxation time scheme~\cite{lallemand2003}, the\
collision operator includes the hydrodynamic moments built into\
independent modes which improves stability.

The incorporation of immersed particles in LB fluid began with Ladd\
et al.~\cite{ladd1994} and Aidun et al.~\cite{aidun1998}. The\
particles were discretized on the lattice as solid nodes immersed in\
fluid nodes, and the lattice was updated between fluid or solid states as the particles translated\
and rotated through the system. In order to resolve the lubrication\
forces that arise between surfaces in close contact,\
a formalism was developed to explicitly add these\
interactions~\cite{nguyen2002}. Recent examples of the discretized\
particle method for ellipsoids include modeling of red blood\
cell dynamics~\cite{janoschek2010} and of colloids at fluid-fluid interfaces~\cite{gunther2013,davies2014}.

LB was first joined with MD in 1999 by Ahlrichs and\
D\"{u}nweg~\cite{ahlrichs1999} by coupling the fluid to the individual\
monomers of a polymer chain. The force of the fluid is imposed upon\
the MD beads using a modified Langevin equation

\begin{equation}
\label{eq:latticeboltzmannfriction}
\overline{F}_{fl}=- \xi_{bead} \left [\overline{v}-\overline{u}(\overline{r},t) \right ] + \overline{f}_s,
\end{equation}

\noindent
where $\xi_{bead}$ is the resistance factor, $\overline{v}$ the\
particle velocity, $\overline{u}(\overline{r},t)$ the fluid velocity\
interpolated at the given position $\overline{r}$ between the nearest\
grid points, and $\overline{f}_s$ a noise term that follows the\
fluctuation dissipation theorem. The bead resistance $\xi_{bead}$\
is a tunable parameter corresponding to friction.

The LB equation can be run in a stochastic or deterministic\
(noise-free) system, depending on the focus of the investigation. LB\
allows hydrodynamic interactions to be included in MD\
simulations at low computational cost. It is inherently parallelizable\
due to the grid representation of fluid populations, and dramatic\
improvement to speed can achieved using GPU\
processing~\cite{lbgpu2012} which is especially suited to mesh systems\
like LB. Boundary conditions are enforced at low cost by assigning\
nodes as ``boundary nodes." The populations streaming to boundary nodes\
are reflected in the next step, known as the ``bounce-back" rule.

\subsection{Colloid particle model}
\label{sec:colloidmodel}
Several models for protein coarse graining currently\
exist, including MARTINI~\cite{monticelli2008},\
shape based coarse graining~\cite{arkhipov2006}, and the models listed previously. A review of methods\
has been given by Baaden~\cite{baaden2013}.

\begin{figure*}
	\centering
	\begin{tabular}{m{2.9cm}  m{5.6cm}  m{3.8cm} }
		\resizebox{1.0\hsize}{!}{\includegraphics{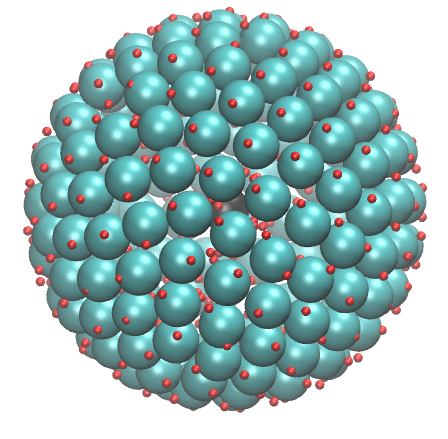}} &	
		\resizebox{1.0\hsize}{!}{\includegraphics{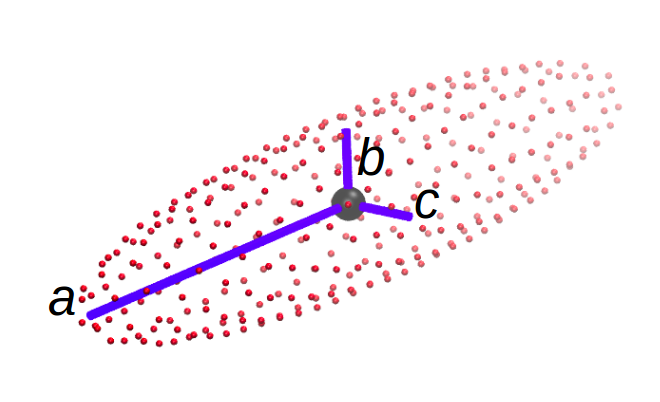}} &	
		\resizebox{1.0\hsize}{!}{\includegraphics{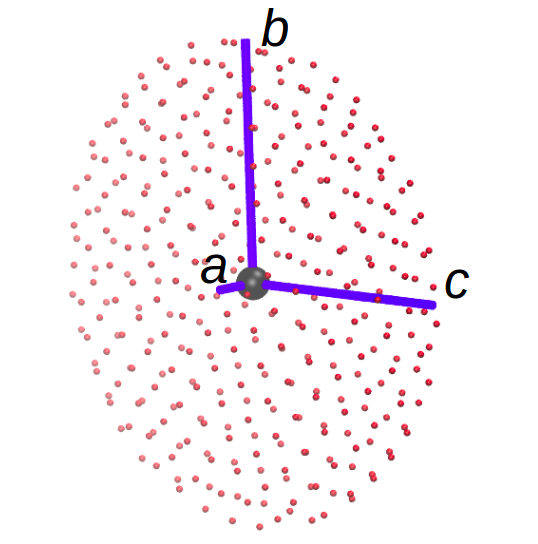}}
	\end{tabular}
	\caption{Raspberry colloid particles. Left: sphere $(a=b=c)$. Middle: prolate ellipsoid $(a>b=c)$. Left: oblate ellipsoid $(a<b=c$) Cyan beads are volume-filling sites (shown only in the sphere) and red beads are hydrodynamic sites only. Blue lines indicate the semi-axes $a$, $b$, and $c$. At the centers of masses are black beads that all other particles in the raspberry are virtualized with respect to.}
	\label{figure:rasp}
\end{figure*}

When hydrodynamics is modeled with LB, the raspberry colloid model~\cite{dunweg2004} is\
a practical solution for simulating coarse-grained bodies in MD. A raspberry is a\
rigid body with an outer shell of MD beads spread evenly over a surface. Each surface bead\
represents an Oseen force in the LB field. Integrating singularity forces\
over the surface of a sphere recovers the Stokes force, similarly to boundary element methods.\
This rectifies the difference between point forces and distributed\
surface forces when coupling LB to MD. The surface beads are virtual\
particles with respect to a bead at the center of mass of the colloid.\
This requires the body to be rigid but allows rotational degrees of\
freedom and reduced computational time, since only the center bead is\
integrated in the Verlet scheme. The raspberry was first introduced by\
Lobaskin and D\"{u}nweg~\cite{dunweg2004} as hollow spheres\
and has since been constructed as a filled body in Fischer et al. and de Graff et al.~\cite{fischer2015rasp2layer1st,degraff2015rasp2layerconfined}.\
The body filling beads are also virtualized with respect to the center\
of mass bead. Figure \ref{figure:rasp} shows different geometries of raspberries.

\section{Methods}
\label{sec:meths}
\subsection{Raspberry construction}
\label{sec:raspconstruction}
Raspberry colloids are constructed in an MD-style scheme using a combination of two forces.\
The first is a harmonic potential between the surface bead and the\
coordinate on the surface with the shortest distance to the bead. The\
second is a Weeks-Chandler-Anderson \cite{wca1971}\
potential that is applied pairwise between the surface beads. When\
integrated in an MD style scheme, the sum of these forces induces the\
beads to spread out evenly over the surface. The spherical raspberry\
colloid particles in this study were constructed using ESPResSo \cite{espresso2013}\
in the same manner as de Graff et al.~\cite{degraff2015rasp2layerconfined}.\
We construct non-spherical raspberry colloid particles here for the first time.\
For a non-spherical body, the construction is more complicated and we\
describe the method in the SI.

All ellipsoids constructed and simulated in this study were ellipsoids\
of revolution, with two degenerate axis lengths $(a,b=c)$.\
Eight prolate $(a > b = c)$ and eight oblate $(a < b = c)$ ellipsoidal raspberry\
colloids of anisotropy $0.1< \phi=\frac{a}{b}<10.0 $ were constructed:\
$\phi = \frac{1}{10}$, $\frac{1}{5}$, $\frac{1}{4}$, $\frac{1}{3}$,\
$\frac{1}{2}$, $\frac{2}{3}$,  $\frac{4}{5}$, $\frac{9}{10}$,\
$\frac{10}{9}$, $\frac{5}{4}$, $\frac{3}{2}$, 2, 4, 5, 10.\
All ellipsoids were of volume $(4/3)\pi (4.25)^{3}$\ nm$^{3}$.\

For all raspberries, two surface shell layers were overlaid. The outer\
serves as the hydrodynamic layer and the inner layer as the\
excluded volume layer. During construction the inner layer beads were\
placed at a radius 0.5~nm less than the radius of the outer layer from the center of the\
colloid. The inner layer was assigned a Lennard-Jones radius of 0.5~nm\
in MD. For ellipsoids the inner layer shell was constructed\
as $(a-0.5,b-0.5,c-0.5)$\ nm. This allowed us to place hydrodynamic\
sites at the same location as the edge of the excluded volume of the\
raspberries. Filling the raspberries with interior hydrodynamic coupling points\
resulted in a model that obeyed Faxen's Law for enhanced drag between parallel plates\
in de Graff et al.~\cite{degraff2015rasp2layerconfined}\
and consequently all raspberries discussed here were filled with\
interior beads.

\subsection{Simulation methods}
\label{sec:simmthds}
All simulations were performed using the MD suite ESPResSo version\
3.2.0, development code version 678-g31c4458~\cite{espresso2013}.\
The hydrodynamic interactions were\
computed using the lattice Boltzmann (LB) method implemented on graphics\
processing units (GPUs)~\cite{lbgpu2012}. Each simulation was performed\
on the Surface machine at Lawrence Livermore National Laboratories on\
eight Intel Xeon E5-2670 processors and one NVIDIA Tesla K40m GPU.\

\subsection{Simulation parameters}
\label{sec:simparams}
A unit system relevant to protein-sized\
colloids particles was established: $M_0=100$~amu, $\sigma=1$~nm,\
$\epsilon=k_{B}298$~K, $t_0=\sigma \sqrt{ \epsilon/M_0 }=6.352$~ps. The integration\
time step was $\Delta t = 0.01~t_0$.

In ESPResSo, the input parameters\
to LB are bead resistance constant $\xi_{bead}$ (eq.~\ref{eq:latticeboltzmannfriction}),\
fluid density $\rho$, lattice spacing\
$a_{grid}$, kinematic fluid viscosity $\nu$ and the fluid time\
step $\Delta t_{LB}$. The density and kinematic viscosity were determined from water at 298~K:\
$\rho = $~997~kg$/$m$^{3} =$~6.004~$M_0/\sigma^3$, $\nu= 8.934\times10^{-7}$~m$^{2}/$s~$=5.677\sigma^2 / t_0$ \cite{sengers1986}.\
The bead resistance was varied as dicussed later. The bead\
resistance and viscosity were scaled such that the actual simulated\
viscosity was $\nu=0.8~ \sigma^2 / t_0$\
in order to increase the simulated time. Transport measurements\
using normal and scaled parameters were compared to validate the low viscosity system. The fluid update time\
was chosen to be $0.02~t_0$, or twice the MD time step.\

At each integration step, the forces and torques from the virtual\
beads are computed and applied to the non-virtual center of mass bead\
in the body-frame defined by the principal axes of the particle. The\
positions of the center of mass bead were updated using the Velocity\
Verlet algorithm. The rotations of the center of mass bead were\
handled using a quaternion. This quaternion $\overline{q}(t)$\
represents a rotation from the intial configuration to the current\
configuration at time $t$. The quaternions were updated using a\
formalism of the Verlocity Verlet algorithm for rotations that\
includes time derivatives for the quaternion~\cite{omelyan1998,martys1999},\
which is effectively an angular velocity. The postions of the virtual\
beads are subsequently updated with respect to the position and\
quaternion of the center of mass bead.

Two types of stochastic transport simulations were performed: Infinite dilution\
and confinement. In the infinite dilution simulations, a single\
raspberry colloid was simulated in lattice Boltzmann fluid. The\
fluid was allowed to warm up over 1000 integration steps. After the\
warm up, recordings were made every 200 integration steps.
The position of the center of mass bead, the position of a reference bead on\
the surface, and a quaternion with respect to initial orientation\
were recorded. Multiple runs were performed for a total of $2 \times 10^{5}$
recordings or $2.541$~$\mu$s. Infinite dilution conditions were determined using finite\
size scaling (see Section~\ref{sec:methslabdt}). The box sizes for all ellipsoids were cubic with 27, 32,\
38, 48, 55, 64, 76, 94~nm long sides. For spheres the boxes were similarly\
scaled relative to the size of the raspberry. The lab-frame translation, lab-frame rotation, body-frame\
translation, and body-frame rotation diffusivity were determined where appropriate.

Transport under confinement was simulated by observing the translation of a colloid within a\
cylindrical pore. Spherical raspberries were of size $R_{rasp}=4.25$~nm. Multiple pore\
sizes were simulated such that the non-dimensional pore size was\
$\lambda=R_{rasp}/R_{pore}=$~0.1, 0.2, 0.3, 0.4, 0.5. The simulation box size in the $x$ and $y$ directions depended\
on the pore size. In the $z$ direction the box\
size was 85~nm. Beads defining the pore surface were positioned\
5~nm apart, centers to centers distance, around the pore.\
The Lennard-Jones radius was $r_p=$ 6.25~nm for the pore beads and $r_r=$ 0.5~nm for the\
volume surface layer of colloid beads. The interaction potential between these particles was\
defined as a purely repulsive WCA potential with $\sigma_{WCA}= r_r + r_p$\
and $\epsilon = k_{B}298$~K. Lattice Boltzmann boundary nodes were positioned one half grid spacing,\
0.5~nm, outside the pore wall in order to create a hydrodynamic barrier\
at the pore wall. After the fluid was warmed, recordings of the position of the center of mass bead\
were taken every 200 integration steps for a total of $1.2 \times 10^{6}$ recordings.\
The motion of the colloid particle was analyzed in the pore-frame\
which will be explained in a later section.

In a third type of simulation, Stokes flow over a raspberry was used to calculate colloid resistance.\
Colloid particles were held fixed in the center of a\
cubic simulation box with length 50~nm with lattice Boltzmann boundary nodes along the\
$+y$ and $-y$ faces of the simulation box with fluid velocity imposed\
as $\overline{v}=[0,0,2.20]$\ m/s at these boundaries. The thermostat was turned off. This set-up\
created a Galilei transform of a particle moving through a quiescent\
fluid in a wide channel with particle Reynolds number\
$Re=0.0135$. After the flow was allowed to develop fully, the force\
on each bead on the raspberry colloid was recorded. The mean of\
100 force measurements was recorded. Finite size scaling was\
performed to ensure that the proximity to the boundary did not affect\
the results. The colloid resistance was determined based on Equation~\ref{eq:stokesflow}:

\begin{equation}
\label{eq:stokesflow}
\sum_{beads} \overline{F}_{bead} = -\xi_{col} \overline{v}
\end{equation}

\noindent
where the force on every bead was added to determine the total\
frictional force experienced by the raspberry. Increasing values of\
of bead resistance were used. This colloid resistance $\xi_{col}$ was\
compared to the resistance determined from stochastic transport\
simulations.
%%%%%%%%%%%%%%%%%%%%%%%%%%%%%%%%
\subsection{Lab-frame translational diffusivity}
\label{sec:methslabdt}
Lab-frame translational diffusivities were determined by the slope of the mean squared\
displacement of position versus lag time:

\begin{equation}
\label{eq:dt_lab}
D_{t}=\lim_{\tau\rightarrow \infty} \frac{MSD}{2d\tau}=\lim_{\tau\rightarrow \infty} \frac{\left<(\overline{r}(t)-\overline{r}(t+\tau))^{2}\right>}{6\tau}
\end{equation}

\noindent
where $\tau$ represents the lag time.

In order to eliminate the hydrodynamic interactions of the colloids\
with their images, finite size scaling was used. The result of Hasimoto~\cite{hasimoto1959}\
of forces on period arrays of spheres due to fluid flow can be used to\
determine diffusivity at infinite dilution:

\begin{equation}
	\label{eq:hasimoto}
	D^{s}(L)=D^{0}-\frac{2.837k_{B}T}{6\rho\eta}\frac{1}{L}
\end{equation}

\noindent
where $D^s(L)$ is the self diffusivity at a certain box size $L$, $D^{0}$ is the\
diffusivity in the given limit of infinite dilution, $\rho$  the density, and\
$\eta$ the dynamic viscosity. Eight\
different simulation sizes were used and the intercept of $D^s$\
versus $L^{-1}$ gave the diffusivity at infinite dilution $D^{0}$.

The colloid resistance to translation $\xi^{t}_{col}$ was determined\
from the Einstein equation

\begin{equation}
	\label{eq:einstein_trans}
	D^{0}_{t}= \frac{k_{B}T}{\xi^{t}_{col}}.
\end{equation}

The colloid resistance to diffusive motion at infinite dilution was\
determined for colloids at $R_{rasp}=3.25$~nm and\
$5.25$~nm at various bead resistance values. A trend line was established\
for $R_{rasp}$ versus $\xi^{t}_{col}$ for spherical raspberry\
particles of size $R_{rasp}=$2.25, 2.75, 3.25, 3.75,\
4.25, 4.75, 5.25, 5.75~nm based on a least squares fit.\
The bead resistance values for these trendlines were 50 and 1000~$M_{0}/t_{0}$.

For ellipsoidal colloids, the trendline was used to find a translational effective\
hydrodynamic radius $R_{eff,t}$ and the geometric factor $A({\phi})$ for resistance due to\
the anisotropy was determined from

\begin{equation}
	\label{eq:perrin_a}
	A({\phi}) = \frac{a}{R_{eff,t}}
\end{equation}

\noindent
and compared to the analytical values by Perrin \cite{perrin1934,perrin1936}.\
The formula for the Perrin factor is given in the SI.\
The error was determined by applying Equation~\ref{eq:hasimoto} to the\
standard deviation of the separate simulation runs at each box size.
%%%%%%%%%%%%%%%%%%%%%%%%%%%%%%%%%%%%%%%%%%%%%
\subsection{Lab-frame rotational diffusivity}
\label{sec:methslabdr}
Lab-frame rotational diffusivities were determined by the slope of the\
mean squared displacement of angle versus lag time:

\begin{equation}
\label{eq:dr_lab}
D_{r}=\lim_{\tau\rightarrow \infty} \frac{MSD}{2d\tau}=\lim_{\tau\rightarrow \infty} \frac{\left<\Delta \Theta(t+\tau)^{2}\right>}{6\tau}
\end{equation}

The angular displacement was determined by measuring the vector\
pointing from the center of mass bead to the reference bead on the surface\
of the raspberry particle. This vector $\overline{p}(t)$ changes direction with the rotation of the\
rigid raspberry body, and the angular displacement $\Delta \Theta$\
is determined from

\begin{equation}
\label{eq:delta_theta}
\Delta \Theta = \cos^{-1} { \left ( \hat{p}(t) \cdot \hat{p} (t+\tau) \right ) }
\end{equation}

\noindent
where $\hat{p}$ is the unit vector of $\overline{p}$. For our box sizes, the periodic image does not affect the\
rotational motion and $D_{r}$ was taken as the average of all runs\
at all box sizes.

The colloid resistance to rotation $\xi^{r}_{col}$ was determined similarly\
to the case of translation and a trendline was established for\
$R^{3}_{col}$ versus $\xi^{r}_{col}$ for the spherical colloids.

For ellipsoidal colloids, the trendline was used to find a rotational effective\
hydrodynamic radius $R_{eff,r}$ and the geometric factor $B({\phi})$ for resistance due to\
the anisotropy was determined from

\begin{equation}
	\label{eq:perrin_b}
	B({\phi}) = \left ( \frac{a}{R_{eff,r}} \right ) ^{3}
\end{equation}

\noindent
and compared to the analytical values by Perrin \cite{perrin1934,perrin1936}.\
The formula for the Perrin factor is given in the SI.
The error was determined from the standard deviation of all\
separate simulation runs at all box sizes.

%%%%%%%%%%%%%%%%%%%%%%%%%%%%%%%%%%%%%%%%%%
\subsection{Body-frame translational diffusivity}
\label{sec:methsbfdt}
An anisotropic particle has anisotropic resistance to motion through a fluid. A prolate ellipsoid with semi-axes $(a > b = c)$ should have\
higher diffusivity along the principal axis $a$ compared to $b$ and $c$.\
An oblate ellipsoid with semi-axes $(a < b = c)$ should have\
lower diffusivity along the principal axis $a$ compared to $b$ and $c$.\
The anisotropic geometry shifts the translational effective hydrodynamic radius\
when the body translates along the individual principal axes of the ellipsoid.\
Figure~\ref{figure:rasp} shows the principal axes of a prolate and an oblate ellipsoid.\
We examined motion in the body-frame of a raspberry colloid\
particle. The principal axes along the semi-axes $a$, $b$, and $c$\
define the body-frame as $x_{b}$, $y_{b}$, and $z_{b}$.
As the colloid particle rotates, the body-frame rotates with it.\
We sought to observe the effects of anisotropy on the body-frame diffusive motion and\
compare the results to the analytical values of Happel and Brenner \cite{happelbrenner1983}.

We followed the method of Han et al.~\cite{han2006} to develop a body-frame trajectory and analyze\
body-frame transport. The body-frame is related to the lab-frame\
by a rotation. Han used a two\
dimensional body-frame rotation to experimentally track ellipsoidal\
particle motion confined to two dimensions.\
We require a three dimensional\
rotation which was made straightforward by the use of quaternions.\
ESPResSo tracked the quaternion that would generate the rotation from the initial\
configuration of the colloid particle to the present orientation. We therefore\
used the inverse quaternion to evaluate motion along the body axes.\

The lab-frame displacement over one measurement interval is

\begin{equation}
\label{eq:labdisp}
\delta \overline{X}(t_n) = \overline{X}(t_n) - \overline{X}(t_{n-1}).
\end{equation}

For an inverse quaternion $\overline{q}^{-1}(t) = (w,-x,-y,-z)$ at\
time $t$ the rotation

\begin{equation}
\label{eq:quat}
\delta \overline{X}^{b}_n = \overline{q}^{-1} \delta \overline{X_n} \overline{q}
\end{equation}

\noindent
gives the body-frame displacement over one measurement interval,\
$\delta \overline{X}^{b}_n$. The\
quaternion rotation is given explicitly in the SI.

The total body-frame displacement for time $t_{n}$ was built from a\
summation over all previous body-frame displacements:

\begin{equation}
\label{eq:bodyframedisplacement}
\overline{X}^{b}(t_{n}) = \sum\limits_{k=1}^n \delta \overline{X}^{b}(t_{k})
\end{equation}

\noindent
and the body-frame displacement between time $t$ and lag time $\tau$ was based on\
equation \ref{eq:bodyframetrajectory}:

\begin{equation}
\label{eq:bodyframetrajectory}
	\Delta \overline{X}^{b}(t+\tau) = \overline{X}^{b}(t+\tau) - \overline{X}^{b}(t).
\end{equation}

Mean squared displacements of the trajectories along the principal axes defined by the semi-axes\
$a$,\ $b$,\ and $c$ were analyzed to determine $D_{t,x^{b}_{i}}$, the\
translational diffusivity along body axis $x^{b}_{i}$, $\overline{X}^{b} \in \{x^{b},y^{b},z^{b}\}$:

\begin{equation}
\label{eq:dt_body}
D_{t,x^{b}_{i}}=\lim_{\tau\rightarrow \infty} \frac{MSD}{2d\tau}=\lim_{\tau\rightarrow \infty} \frac{\left<(\Delta x^{b}_{i}(t+\tau))^{2}\right>}{2\tau}.
\end{equation}

The trendline of $\xi^{t}_{col}$ versus $R_{rasp}$ was used to find a translational effective\
hydrodynamic radius $R_{eff,x^{b}_{i},t}$ for motion along the $x^{b}_{i}$\
body axis. The relationship between effective size and\
actual size of the colloid particle is shown as a geometric factor.

For prolate ellipsoids, the geometric factor $E({\phi})$ on resistance to motion along the long singular ($a$) axis was determined from

\begin{equation}
	\label{eq:happel_e}
	E({\phi}) =  \frac{R_{eff,x^{b},t}}{c}
\end{equation}

\noindent
and the geometric factor $F({\phi})$ on resistance to motion along the short degenerate ($b$,\ $c$) axes was determined from

\begin{equation}
	\label{eq:happel_f}
	F({\phi}) =  \frac{R_{eff,y^{b}z^{b},t}}{c}  .
\end{equation}

For oblate ellipsoids, the geometric factor $G({\phi})$ on resistance to motion along the short singular axis ($a$) was determined from

\begin{equation}
	\label{eq:happel_g}
	G({\phi}) =  \frac{R_{eff,x^{b},t}}{c}
\end{equation}

\noindent
and the geometric factor $H({\phi})$ on resistance to motion along the long degenerate axes ($b$,\ $c$) was determined from

\begin{equation}
	\label{eq:happel_H}
	H({\phi}) =  \frac{R_{eff,y^{b}z^{b},t}}{c}  .
\end{equation}

$E({\phi})$,\ $F({\phi})$,\ $G({\phi})$,\ and $H({\phi})$ were compared to the analytical values\
calculated by Happel and Brenner~\cite{happelbrenner1983} which are given in the SI. It should\
be noted that the body frame geometric factors in this section and Section~\ref{sec:methsbfrot}\
are reciprocal to the factors in Section~\ref{sec:methslabdr} and are\
given with respect to $c$ as opposed to $a$.

%%%%%%%%%%%%%%%%%%%%%%%%%%%%%%%%%%%%%%%%%%%%%%%%%%
\subsection{Body-frame rotational diffusivity}
\label{sec:methsbfrot}
The body-frame rotational diffusivity was measured for ellipsoidal\
colloids. A body-frame rotation about the body-frame axis $x^{b}_{i}$\
between time $t-1$ and time $t$ is defined\
as the two dimensional projection of the three dimensional\
lab-frame rotation between time $t-1$ and time $t$ onto the plane perpendicular\
to $x^{b}_{i}$ at time $t-1$. For example, rotation about the\
$x^{b}$ axis, or the rotation about the singular axis $a$, is\
determined by projecting the full rotation onto the $(y^{b}z^{b})_{t-1}$ plane.\
The method for determining body-frame angular displacement can be found\
in the SI.

The total body-frame angular displacement about the $x^{b}_{i}$ body axis for time $n$ is built from a\
summation over all previous body-frame displacements:

\begin{equation}
	\label{eq:bodyframerotdisplacement}
	\Theta_{x^{b}_{i}}(t_{n}) = \sum\limits_{k=1}^n \delta \Theta_{x^{b}_{i}}(t_{k})
\end{equation}

\noindent
and the body-frame displacement between time $t$ and lag time $\tau$ was based on\
equation \ref{eq:bodyframetrajectory}:

\begin{equation}
	\label{eq:bodyframerottrajectory}
	\Delta \Theta_{x^{b}_{i}}(t+\tau) = \Theta_{x^{b}_{i}}(t+\tau) - \Theta_{x^{b}_{i}}(t).
\end{equation}

Mean squared displacements of the trajectories for the semi-axes\
$a$,\ $b$,\ and $c$ were analyzed to determine the $D_{r,x^{b}_{i}}$, the\
rotational diffusivity about body axis $x^{b}_{i}$.

\begin{equation}
	\label{eq:dr_body}
	D_{r,x^{b}_{i}}=\lim_{\tau\rightarrow \infty} \frac{MSD^{2}}{2d\tau}=\lim_{\tau\rightarrow \infty} \frac{\left<(\Delta \Theta_{x^{b}_{i}}(t+\tau))^{2}\right>}{(3/2)2\tau}.
\end{equation}

The $(3/2)$ factor in the denominator of Equation \ref{eq:dr_body}\
was required for lab-frame $D_{r}$ and body-frame $D_{r}$ values to\
agree and is a result of projecting three dimensional motion into\
two dimensions.

The trendline of $\xi^{r}_{col}$ versus $R^{3}$ was used to find a rotational effective\
hydrodynamic radius $R_{eff,x^{b}_{i},r}$ for motion about the $x^{b}_{i}$\
body axis. For prolate ellipsoids, the geometric factor $I({\phi})$ on resistance to motion about the singular ($a$) axis was determined from

\begin{equation}
	\label{eq:dr_bf_i}
	I({\phi}) =  \left ( \frac{R_{eff,x^{b},r}}{c} \right )^{3}
\end{equation}

\noindent
and the geometric factor $J({\phi})$ on resistance to motion about the degenerate ($b$,\ $c$) axes was determined from

\begin{equation}
	\label{eq:dr_bf_j}
	J({\phi}) =  \left ( \frac{R_{eff,y^{b}z^{b},r}}{c} \right )^{3}.
\end{equation}

For oblate ellipsoids, the geometric factor $K({\phi})$ on resistance to motion about the singular axis was determined from

\begin{equation}
	\label{eq:dr_bf_k}
	K({\phi}) =  \left ( \frac{R_{eff,x^{b},r}}{c} \right )^{3}
\end{equation}

\noindent
and the geometric factor $L({\phi})$ on resistance to motion about the degenerate axes was determined from

\begin{equation}
	\label{eq:dr_bf_l}
	L({\phi}) =  \left ( \frac{R_{eff,y^{b}z^{b},r}}{c} \right )^{3}.
\end{equation}

%%%%%%%%%%%%%%%%%%%%%%%%%%%%%%%%%%%%%%%%%
\subsection{Colloid transport under confinement inside cylindrical pores}
\label{sec:methsenhdrag}

\begin{figure}
	\centering
	\resizebox{0.5\hsize}{!}{\includegraphics{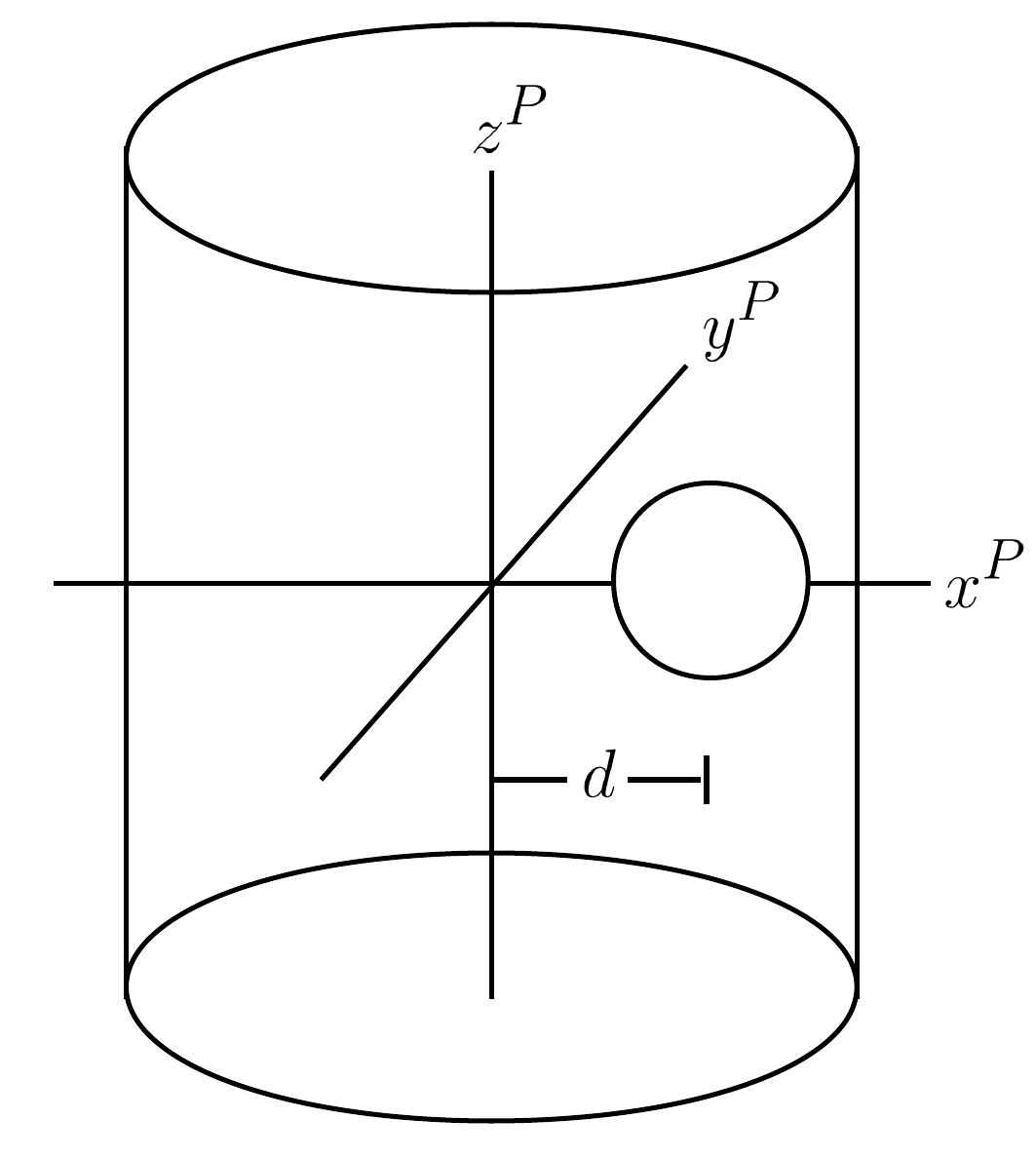}}
	\caption{Pore-frame coordinate system, adapted from \cite{higdon1995}.}
	\label{figure:pore_coords}
\end{figure}

The analysis of transport of\
the colloid in a cylindrical pore was performed by a transformation\
from motion in the lab-frame into motion in the\
pore-frame $\overline{X}^{p} \in \{x^{p},y^{p},z^{p}\}$. As the colloid\
translates within the pore, the pore-frame rotates such that the\
$x^{p}$ axis is equivalent to the radial coordinate, $z^{p}$ is\
equivalent to the axial coordinate, and $y^{p}$ is orthogonal\
to $x^{p}$ and $z^{p}$. The axial coordinate $z^{p}$ requires no\
conversion. The pore coordinate frame is illustrated in Figure~\ref{figure:pore_coords}.

The method for transformation into the pore-frame $\overline{X}^{p}$\
is given in the SI. The total pore-frame displacement about the\
$x^{p}_{i}$ pore-frame axis for time $t_{n}$ is built from a\
summation over all previous displacements:

\begin{equation}
\label{eq:poreframedisplacement}
\overline{X}^{p}(t_{n}) = \sum\limits_{k=1}^n \delta \overline{X}^{p}(t_{n})
\end{equation}

\noindent
and the body-frame displacement between time $t$ and lag time $\tau$ was based on\
equation~\ref{eq:bodyframetrajectory}:

\begin{equation}
\label{eq:poreframetrajectory}
	\Delta \overline{X}^{p}(t+\tau) = \overline{X}^{p}(t+\tau) - \overline{X}^{p}(t_{n})(t) .
\end{equation}

The total pore frame displacement contains the full trajectory in the\
 moving pore-frame, where the $x^{p}$ direction is always the line\
 between the radial center, the colloid particle, and the wall. The\
 $y^{p}$ direction is always normal to this and the axial $z^{p}$\
 direction, and each displacement is analyzed with respect to the\
 pore-frame. Each frame in the trajectory was recorded according to\
 its radial coordinate and a subtrajectory starting with this frame\
 was built from the subsequent displacements of the colloid in the\
 pore-frame. The mean squared displacement for each bin was built\
 from all the subtrajectories of that bin and the slope of the mean\
 squared displacement was used to determine the pore frame\
 translational diffusivity at that bin. The lowest twenty-five lag\
 times in the linear regime of the mean squared displacement for\
 each bin were used for the diffusivity calculation. The pore-frame\
 diffusivity values were normalized by the unbound lab-frame\
 diffusivity at the respective friction to give enhanced drag\
 to finally be compared with enhanced drag data from the literature.

Higdon and Muldowney~\cite{higdon1995} calculated the enhanced drag of a sphere in cylindrical\
pores at off-center nondimensional radial coordinates and nondimensional\
pore sizes using a spectral boundary\
element method. They defined enhanced drag $S_{x^{p}_{i}}$ as\

\begin{equation}
	\label{eq:enhanced_drag_force}
	S_{x^{p}_{i}} = \frac{F_{x^{p}_{i}}}{F_{\infty}}
\end{equation}

\noindent
where $F_{x^{p}_{i}}$ is the force of the fluid in response to motion\
along the  $x^{p}_{i}$ direction in the pore and $F_{\infty}$ is the\
force of the fluid in response to motion in an unbounded system.\
In the present study enhanced drag is equivalently defined as

\begin{equation}
	\label{eq:enhanced_drag_diffusion}
	S_{x^{p}_{i}} = \frac{D_{0}}{D_{x^{p}_{i}}}
\end{equation}

\noindent
where $D_{x^{p}_{i}}$ is the diffusivity of the particle along the\
$x^{p}_{i}$ direction in the pore-frame and $D_{0}$ is the\
diffusivity of the particle at infinite dilution. The error was\
determined from the standard deviation of the mean squared displacement\
normalized by the square root number of independent measurements at each bin,\
making the error bars invisible on the plots.

%%%%%%%%%%%%%%%%%%%%%%%%%%%%%%%%%%%%%%%%%%%%%%%%%%%%%%%%%%%
\section{Results}
\label{sec:rslts}

\subsection{Bead resistance}
\label{rsltsfriction}
\begin{figure}
	\centering
	\begin{tabular}{c c}
		\resizebox{0.48\hsize}{!}{ \includegraphics{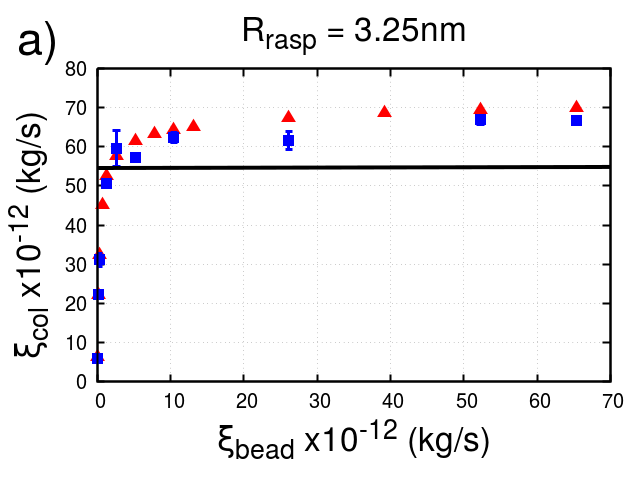} } &
		\resizebox{0.48\hsize}{!}{\includegraphics{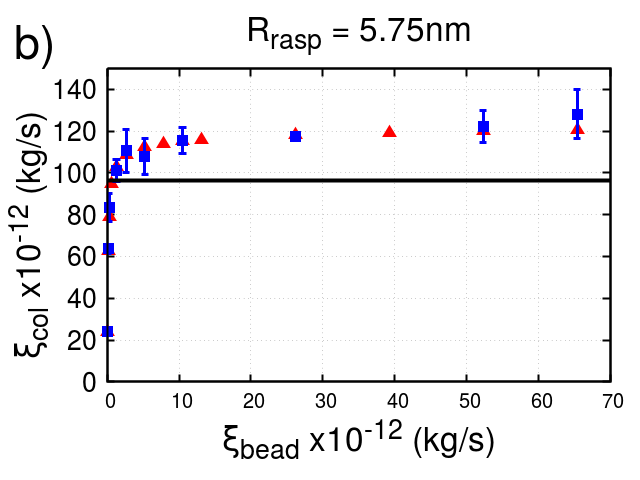} }\\
	\end{tabular}
	\caption{Raspberry colloid translational resistance as a function of bead resistance from Stokes flow and lab-frame translational diffusion results. a) $R=3.25$\ nm. b) $R=5.75$\ nm. Red triangles: colloid resistance from Stokes flow simulation. Blue square: colloid resistance from translational diffusion.}
	\label{figure:friction_trans}
\end{figure}

The input friction value for bead resistance in Equation~\ref{eq:latticeboltzmannfriction}\
has an empirical effect on the relationship between MD beads and LB fluid~\cite{dunweg2008}.\
Figure~\ref{figure:friction_trans} shows the translational resistance of the colloid particle determined in the Stokes\
flow simulations and in the infinite dilution simulations. These\
simulations relate the bead resistance from Equation~\ref{eq:latticeboltzmannfriction}\
to the colloid resistance.\
The colloid resistance $\xi^{t}_{col}$ is determined from Equation~\ref{eq:einstein_trans} for the diffusion simulations and from Equation~\ref{eq:stokesflow} for the Stokes flow simulations.

As the bead resistance\
is increased, the colloid translational resistance increases\
steeply initially but then reaches a maximum value around $\xi_{bead}=1000M_{0}/t_{0}=26.14\times 10^{-12}$kg$/$s. The raspberry allows\
LB fluid to pass through because each MD bead\
is only a point particle to LB. The raspberry becomes more water-tight\
to LB fluid as the bead resistance increases.\
Becoming completely water-tight and having no-slip boundary conditions should result in a colloid resistance\
equal to the Stokes result $\xi^{t}_{col}=6\pi\eta R$. Figure~\ref{figure:friction_trans} shows the\
limiting resistance value for the raspberries\
to be $28.3\%$ and $25.3\%$ higher than the Stokes result for $R_{rasp}$=3.25nm and $R_{rasp}$=5.75nm, respectively.

\begin{figure}
	\centering
	\begin{tabular}{c c}
        \resizebox{0.48\hsize}{!}{ \includegraphics{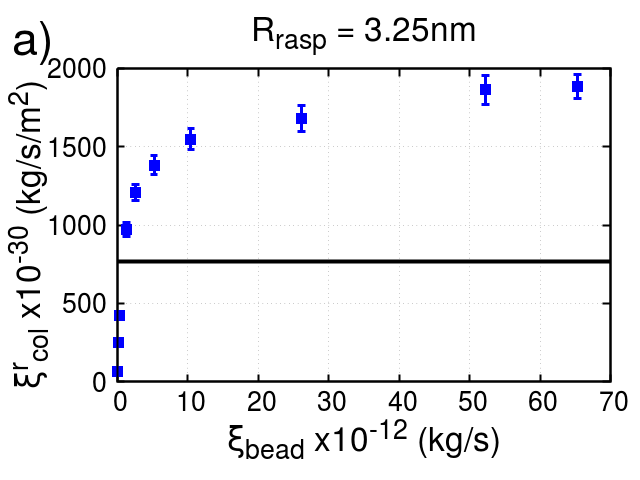}} &
        \resizebox{0.48\hsize}{!}{ \includegraphics{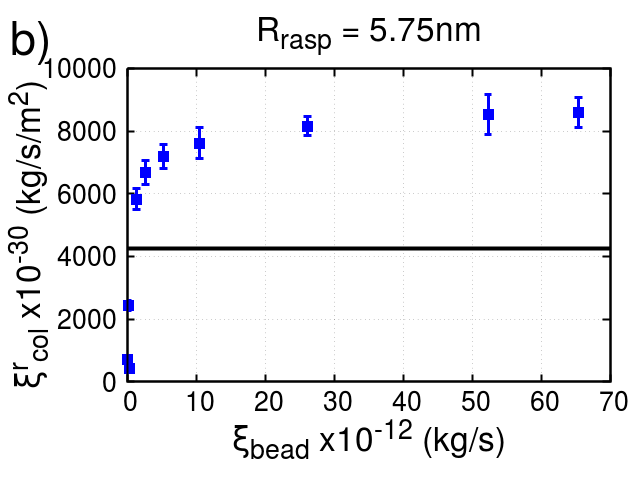} }\\
	\end{tabular}
	\caption{Colloid rotational resistance as a function of bead resistance from lab-frame rotational diffusion results on raspberry colloid particles. a) $R=3.25$~nm. b) $R=5.75$~nm. }
	\label{figure:friction_rot}	
\end{figure}

Figure~\ref{figure:friction_rot} shows the rotational resistance of the colloid particle\
determined from the diffusion. The\
colloid rotational resistance approaches a limiting value, but the input\
bead resistance required is higher for colloid rotational resistance than colloid translational resistance.\
Larger bead resistance values were not allowed due to numerical instabilities\
in the simulations. The maximum value obtained for colloid rotational resistance was $146\%$ and $102\%$ greater than the\
Stokes-Einstein-Debye result $\xi^{r}_{col}=8\pi\eta R^{3}$ for $R_{rasp}$=3.25~nm and $R_{rasp}$=5.75~nm, respectively.

The bead resistance $\xi_{bead}=$50 $M_{0}/t_{0}$ resulted in translational\
colloid resistance values close to the Stokes result for the two\
spherical raspberries of different size. Transport behavior may also depend on the\
leakiness of the colloid, therefore the three resistance values under\
investigation are $\xi_{bead}=$50, 200, 1000 $M_{0}/t_{0}$\
which we call low, mid and high bead resistance.

%%%%%%%%%%%%%%%%%%%%%%%%%%%%%%%%%%%%%%%%%%%%%%%%%%%%%%%%%%%%%%%%%
\subsection{Spheres: Radius}
\label{sec:rsltsradius}
\begin{figure}
	\centering
	\begin{tabular}{c c}
		\resizebox{0.48\hsize}{!}{ \includegraphics{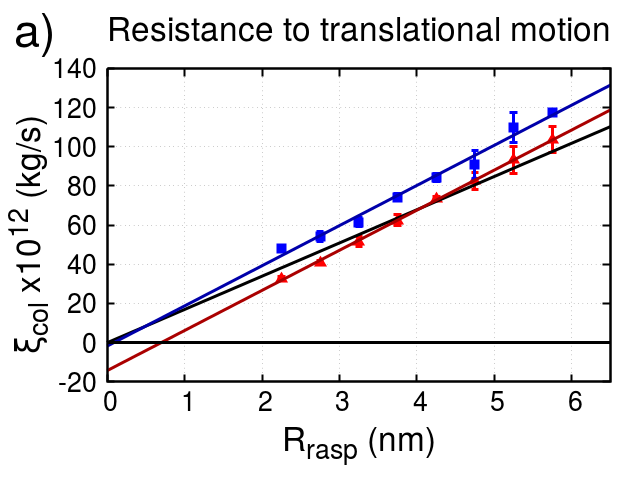} } &
		\resizebox{0.48\hsize}{!}{ \includegraphics{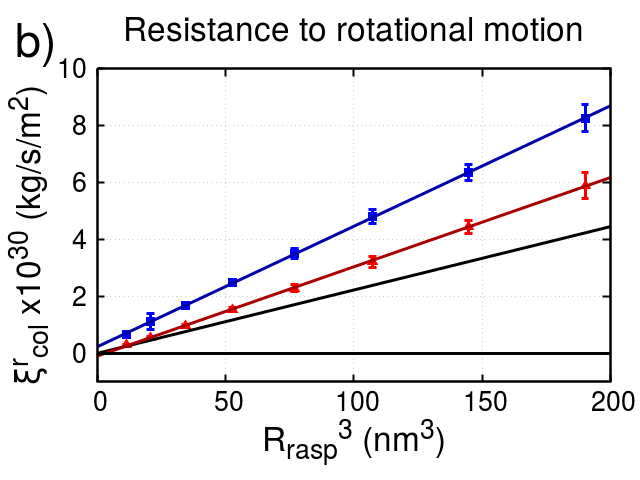} }\\
	\end{tabular}
	\caption{Colloid translational (a) and rotational (b) resistance as a function of colloid radius. Red triangles represent the results of low bead resistance, blue squares represent high bead resistance. The black lines represent the Stokes-Einstein result for translational resistance and the Stokes-Einstein-Debye result for rotational resistance, respectively, and red and blue lines represent the trendlines for low and high bead resistance, respectively.}
	\label{figure:radius}
\end{figure}

The relationships between colloid radius and translational and rotational\
resistance were established for two resistance values: $\xi_{bead}=$ 50 and 1000 $M_{0}/t_{0}$.\
Figure~\ref{figure:radius} presents these data. All raspberry particles\
show linear relations between $\xi^{t}_{col}$ and $R_{rasp}$ as well as between\
$\xi^{r}_{col}$ and $R_{rasp}^{3}$.

\begin{figure}
	\centering
	\begin{tabular}{c c}
		\resizebox{0.48\hsize}{!}{ \includegraphics{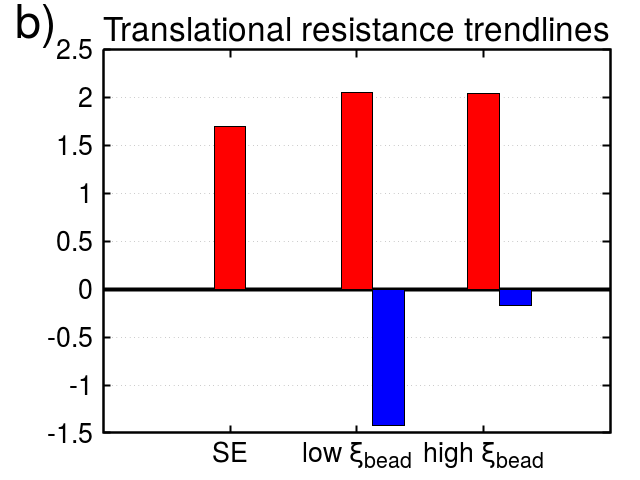} } &
		\resizebox{0.48\hsize}{!}{ \includegraphics{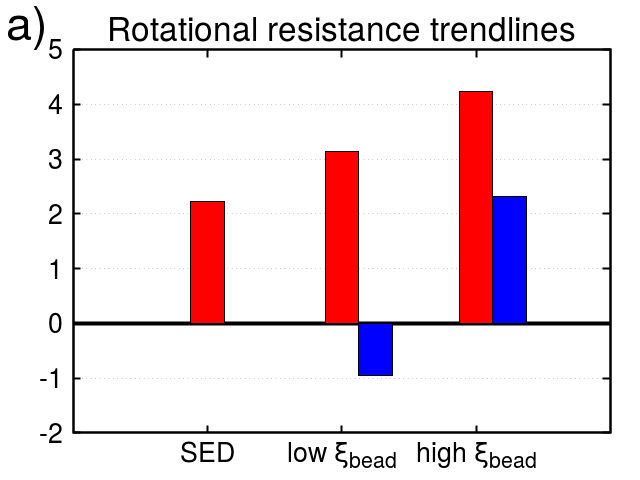} }\\
	\end{tabular}
	\caption{Fits for colloid resistance to translational and rotational lab-frame transport trends. Stokes-Einstein and Stokes-Einstein-Debye with simulation results from different friction values. a: translation; Red bar: slope $\times 10^{2}$; Blue bar: intercept $\times 10^{12}$. b: rotation; Red bar: slope $\times 10^{2}$; Blue bar: intercept $\times 10^{29}$. Units of slopes are kg~s$^{-1}$m$^{-1}$, units of intercepts are kg~s$^{-1}$ for translation, kg~s$^{-1}$~m$^{-2}$ for rotation.}
	\label{figure:trendlines}
\end{figure}

Figure~\ref{figure:trendlines} gives a bar chart representation of\
the least squares fit equations of the lines in Figure~\ref{figure:radius}\
between $\xi^{t}_{col}$ and $R_{rasp}$ and between $\xi^{r}_{col}$ and\
$R_{rasp}^{3}$. The theoretical slopes are $6 \pi \eta R_{rasp}$ for\
translation and $8 \pi \eta R^{3}_{rasp}$ for rotation. The slope was\
greater than the theoretical values for\
each configuration and did not pass through the origin. For\
translation, the slopes of both bead resistanes were similar, and the\
intercept for low resistance was an order of magnitude larger compared to high resistance.\
For rotation, the slope of high bead resistance was 24.2\% higher than\
low bead resistance, and the intercepts were within an order of magnitude.

The nature of LB leads to a renormalization of the hydrodynamic radius\
of MD beads in LB~\cite{ahlrichs1999,dunweg2008}.\
With raspberry particles, the designed radius and hydrodynamic radius are offset as shown by these data and explored by Fischer et al.~\cite{fischer2015rasp2layer1st}.\
Within the renormalized system, however, the\
raspberry particles are defined by a transport frame that is\
self-consisent. Later on, the data will show that the resistance of\
nonspherical particles were also consistent within this frame.\

The equations for the trend lines between $\xi^{t}_{col}$ and\
$R_{rasp}$ and between $\xi^{r}_{col}$ and $R_{rasp}^{3}$ for low\
bead resistance and high bead resistance are given in Table 1 of\
the SI. If the methods and parameters in this study are repeated,\
these trendlines will be useful as a systematic calibration between\
designed radius (including effective radii for nonspherical bodies)\
and hydrodynamic radius.

%%%%%%%%%%%%%%%%%%%%%%%%%%%%%%%%%%%%%%%%%%%%%%%%%%%%%%%%%%%%%%%%
\subsection{Spherical colloid transport under confinement}
\label{sec:rsltsenhdrag}
Enhanced drag on spherical colloids due to confinement in cylindrical\
pores was simulated and analyzed in the pore-frame. Enhanced drag values\
depend on direction within the pore-frame, radial coordinate, and\
relative colloid size. The nondimensional radial coordinate is defined as

\begin{equation}
	\label{eq:beta_definition}
	\beta=\frac{d}{R_{pore}-R_{rasp}}
\end{equation}

\noindent
where $d$ is the distance from the center of the pore to the center of\
the raspberry. The nondimensional colloid size is defined as

\begin{equation}
\label{eq:lambda_definition}
\lambda=R_{rasp}/R_{pore}.
\end{equation}

Figure \ref{figure:enhanced_drag} shows the enhanced drag in the $x^{p}$, $y^{p}$, and\
$z^{p}$ directions. The results of simulations for each pore size are shown\
in the same color as points connected by solid lines across the $\beta$
axis. The dotted lines are the enhanced drag results of Higdon and\
Muldowney~\cite{higdon1995}.

We find high values of $D_x$ at small\
values of $\beta$, where the colloid is close to the pore center. This\
is because motion is biased to the $x^{p}$ direction close to the\
pore center. The result is that enhanced drag in the $x^{p}$ direction is too low\
and the enhanced drag in the $y^{p}$ direction is too high.

Enhanced drag generally increases as $\lambda$ and $\beta$ increase in the simulations.\
For the $x^{p}$ direction of motion, the colloid particles\
approach the pore wall head-on. The $x^{p}$ data\
are shown in Figure~\ref{figure:enhanced_drag}.a, \ref{figure:enhanced_drag}.d, \ref{figure:enhanced_drag}.g. The\
transport of the raspberries shows close agreement with the\
enhanced drag values of Higdon and Muldowney.\
As bead resistance increased the enhanced drag data of Higdon and\
Muldowney was increasingly better reproduced. The enhanced drag increased up to large values of $\beta$,\
where the colloid was so close to the pore wall that there were few\
fluid lattice nodes between the raspberry surface and boundary nodes. Enhanced drag\
tracked most closely to the reference values in the $y^{p}$\
direction (Figure~\ref{figure:enhanced_drag}.b, \ref{figure:enhanced_drag}.e, \ref{figure:enhanced_drag}.h) with high bead resistance, as well.

For the $z^{p}$ axial direction of motion (Figure~\ref{figure:enhanced_drag}.c, \ref{figure:enhanced_drag}.f, \ref{figure:enhanced_drag}.i) resistance was increased for\
larger colloid to pore ratios. As bead resistance increases the enhanced drag increases,\
which is consistent with the $x^{p}$ and $y^{p}$ directions.

The enhanced drag along the $x^{p}$ coordinate, $S_{z}$, is lower\
than Higdon and Muldowney's calculations for small pores or $\lambda > 0.2$.\
The minima of the reference values for $S_{z}$ are at off-center\
coordinates. The pressure force is maximum at the center\
and minimum at the pore wall, the lubrication force is minimum at the\
center and maximum at the pore wall, and the sum is minimized\
at an off-center position. This phenomenon is similar to the well-known\
Segr\'{e}-Silberberg effect\cite{segre1961}, in which lateral\
($x^{p}$ direction) lift forces move particles into off-center\
positions but here the particle and fluid are not at the same\
velocity. As the pore size decreases, the location\
$\beta$ and depth of the minimum in $S_{z}$ increases. This behavior\
was not captured by the simulations.

We also ran pore confinement simulations with raspberries that did\
not have the extra hydrodynamic-only layer, in order to show the\
importance of the extra hydrodynamic coupling points in reproducing\
enhanced drag. Since the two-layer filled raspberry colloid particles\
showed the highest agreement with the spectral boundary element data\
of Higdon and Muldowney, these single layer filled raspberry\
simulations were conducted at high friction coupling values. These\
confinement data are presented in Figure 4 of the SI and are lower\
in enhanced drag than the two layer filled raspberry particles. This\
supports our claim that raspberries particles should have outer\
surface beads at the design radius.

The resistance value under further investigation is $\xi_{bead}=$ 1000 $M_{0}/t_{0}$.\
Based on the enhanced drag data the raspberries are least leaky to LB at this high resistance value.\
Within the renormalized frame described in Section ~\ref{sec:rsltsradius},\
the high bead resistance gives consistent transport behavior exhibiting\
enhanced drag within cylindrical pores.

%-eps-converted-to.pdf
\begin{figure*}
	\begin{tabular}{|c c c|}
		\hline
		\resizebox{0.3\hsize}{!}{ \includegraphics{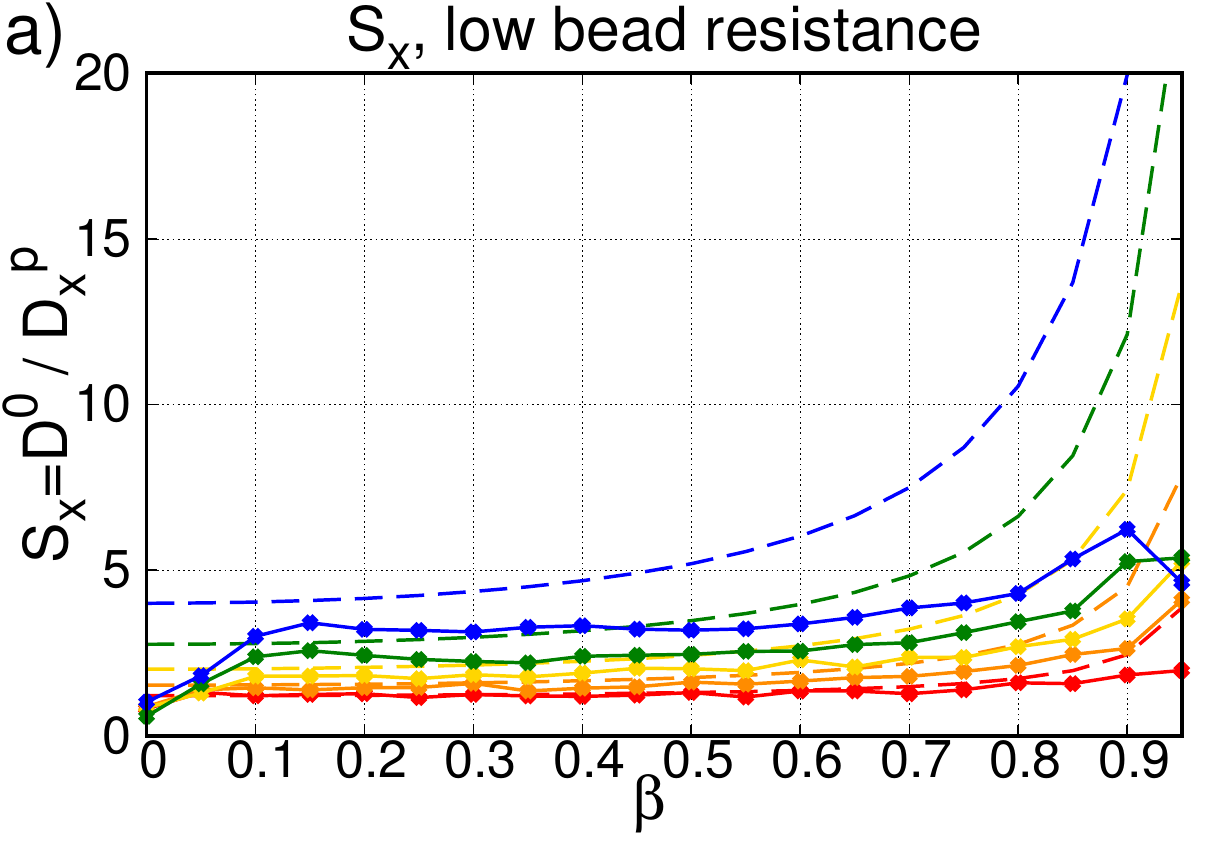}} &
		\resizebox{0.3\hsize}{!}{ \includegraphics{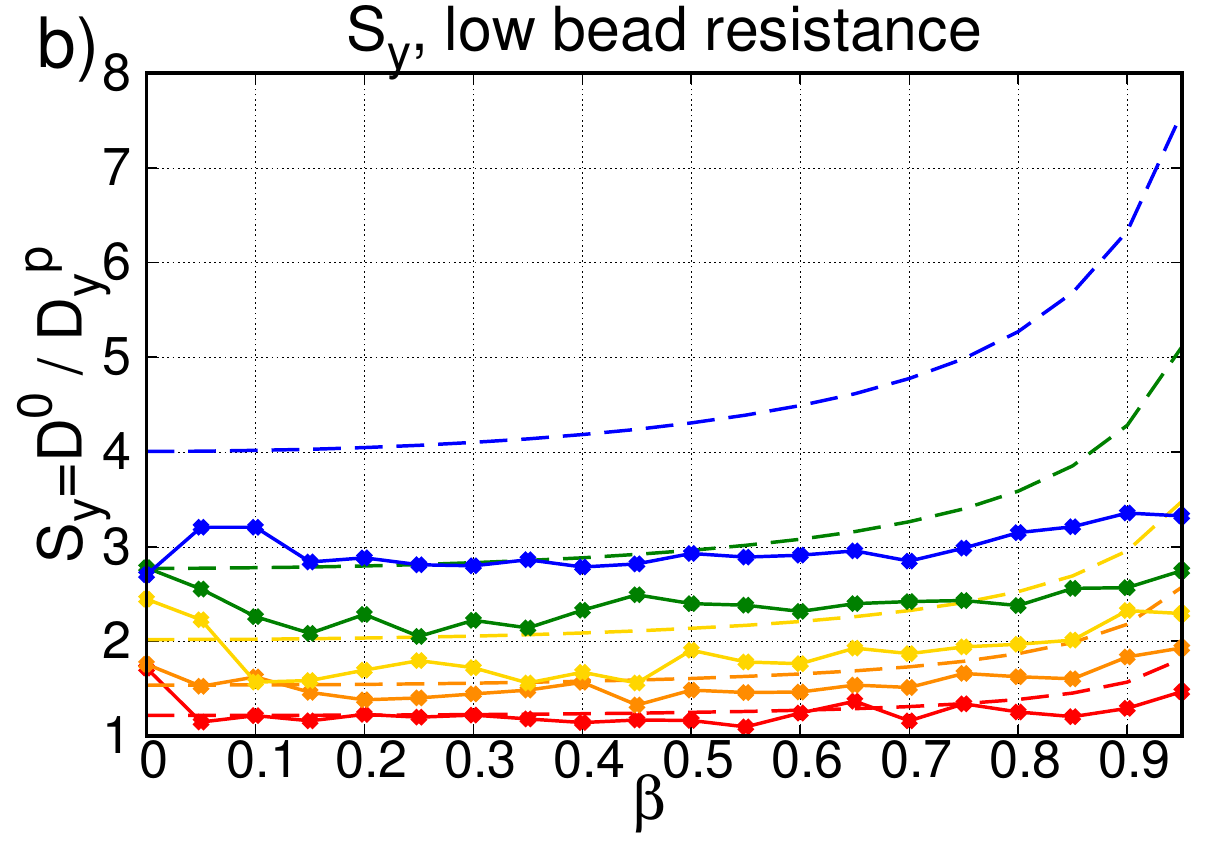}} &
		\resizebox{0.3\hsize}{!}{ \includegraphics{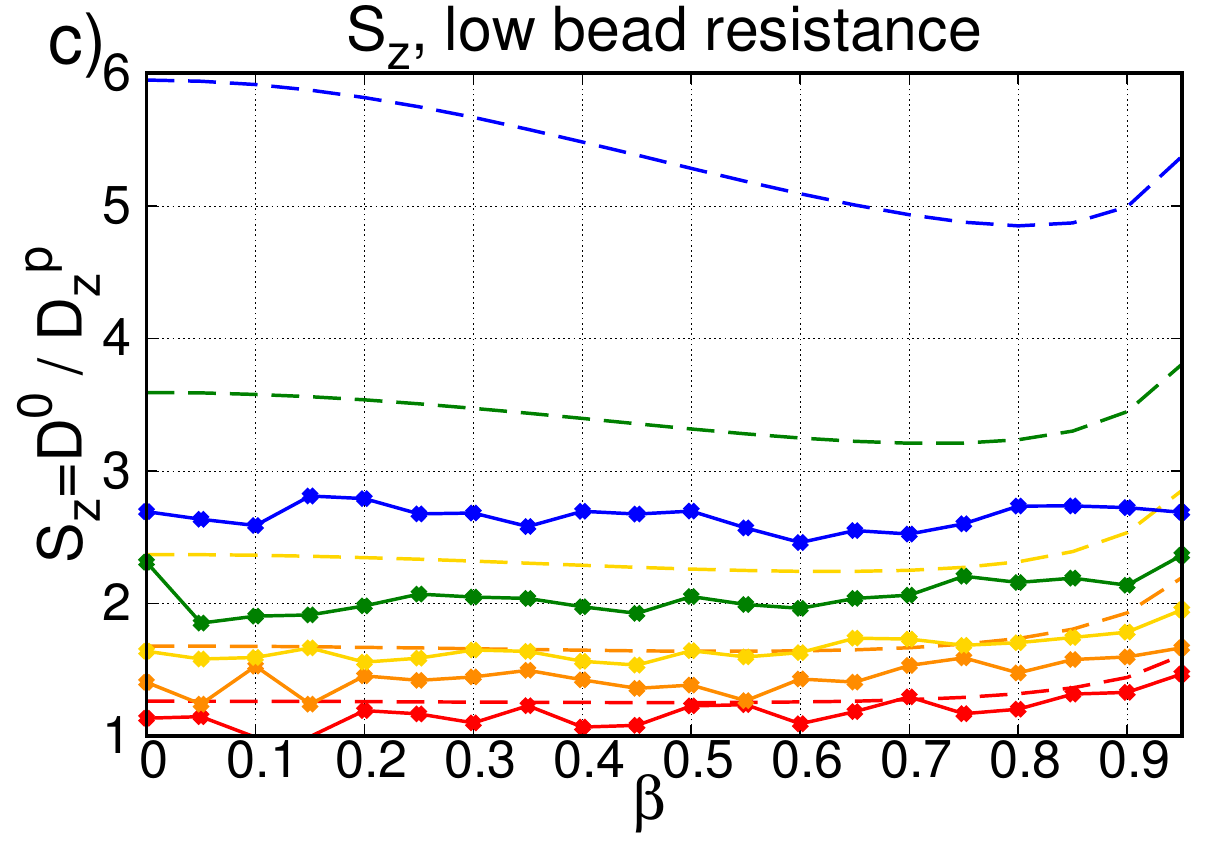}}\\
		\resizebox{0.3\hsize}{!}{ \includegraphics{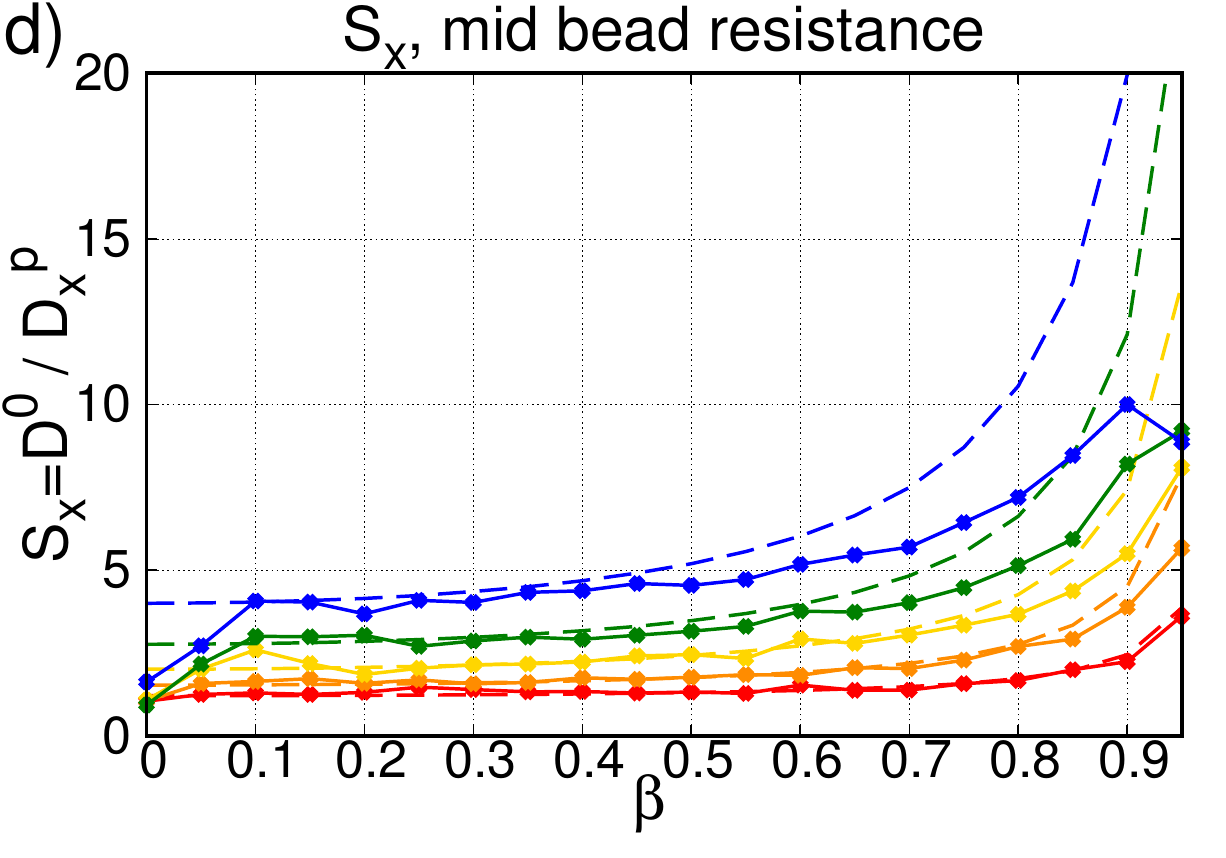}} &
		\resizebox{0.3\hsize}{!}{ \includegraphics{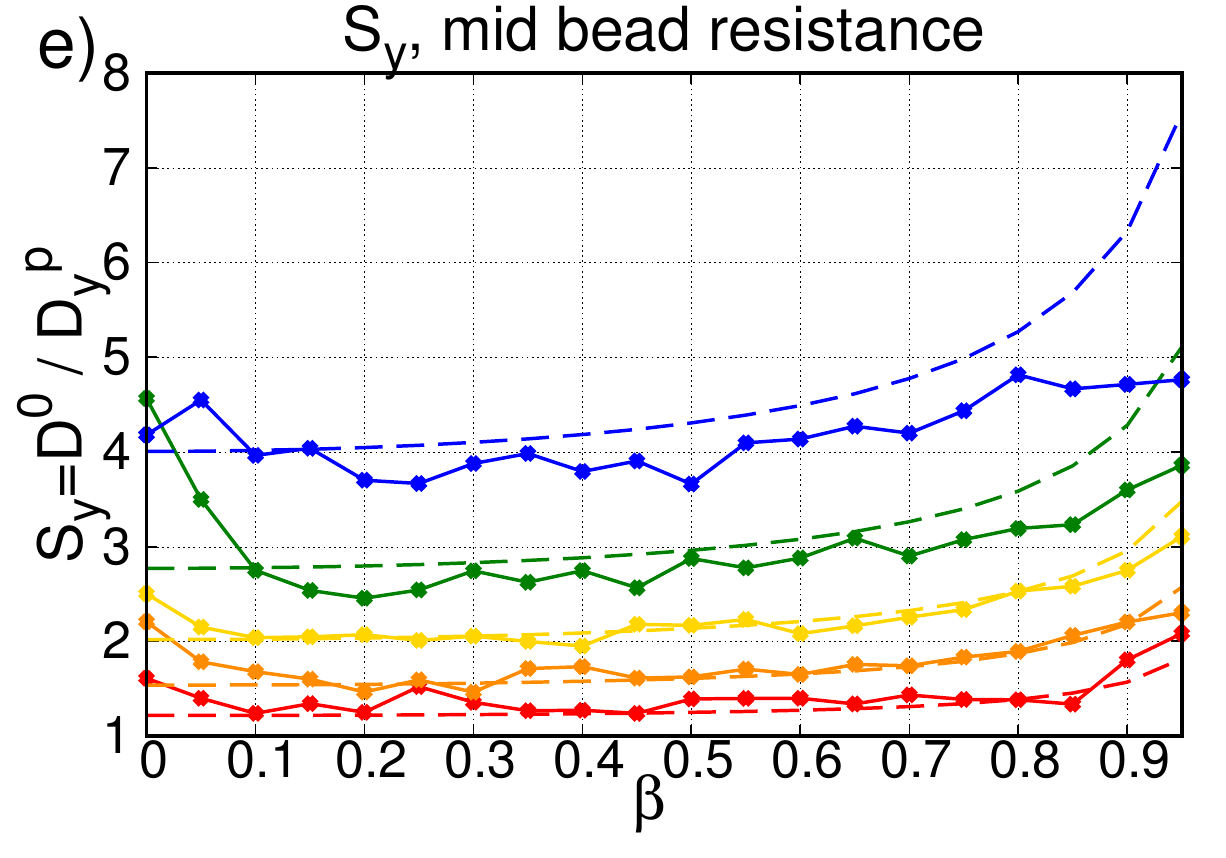}} &
		\resizebox{0.3\hsize}{!}{\includegraphics{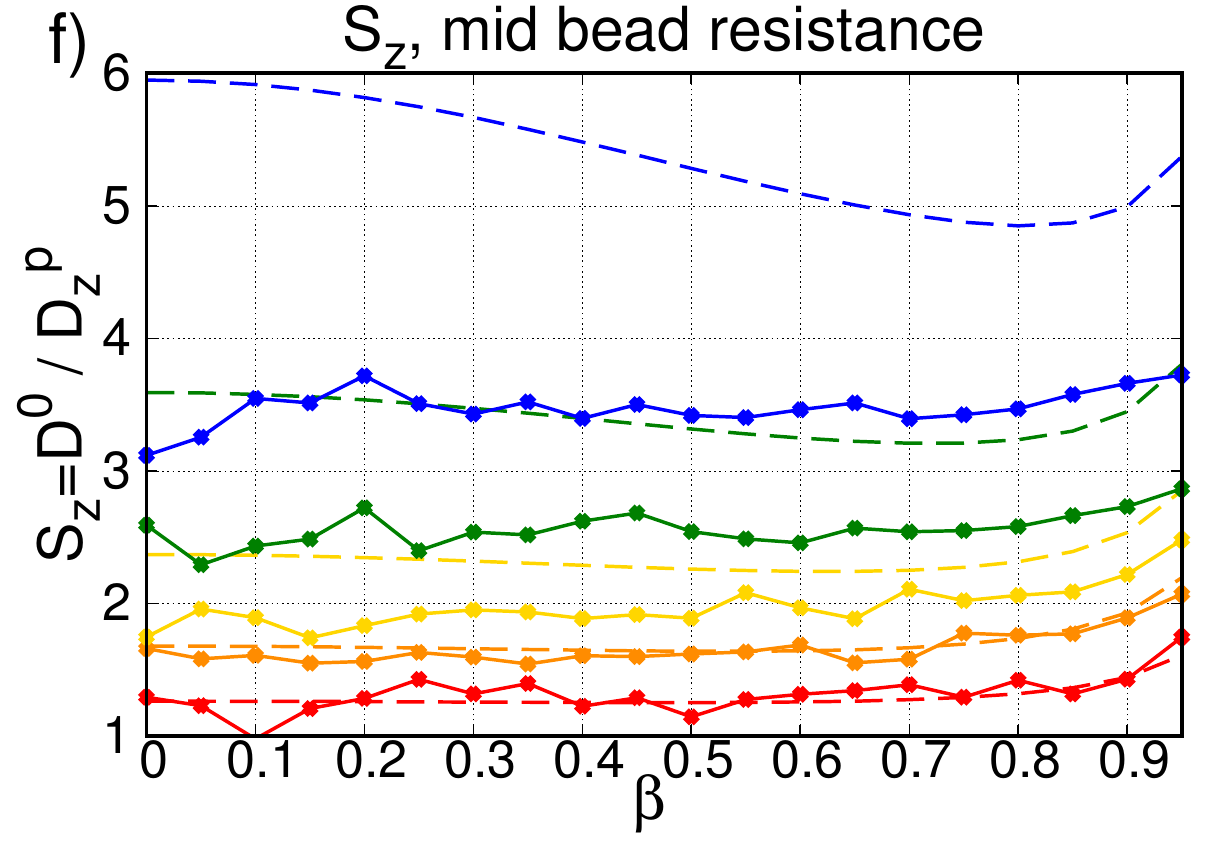}}\\
		\resizebox{0.3\hsize}{!}{\includegraphics{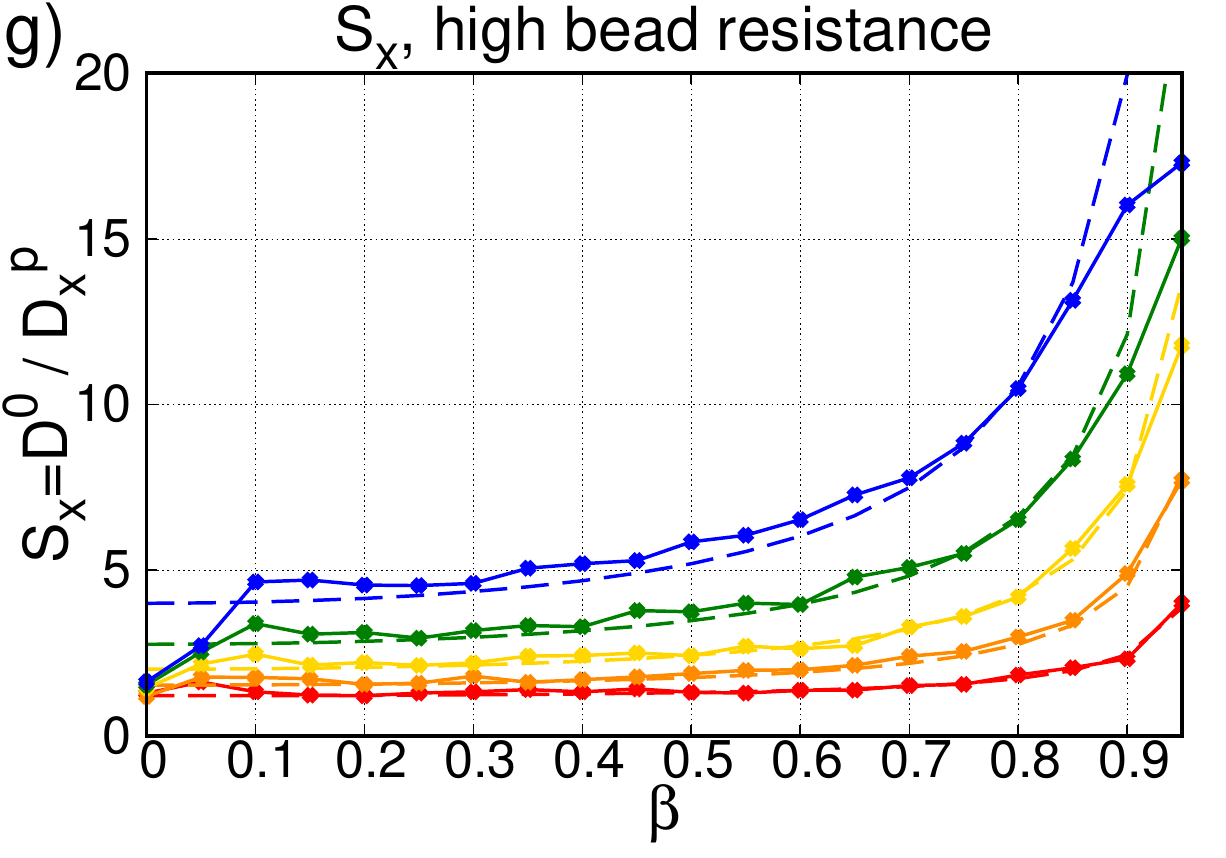}} &
		\resizebox{0.3\hsize}{!}{\includegraphics{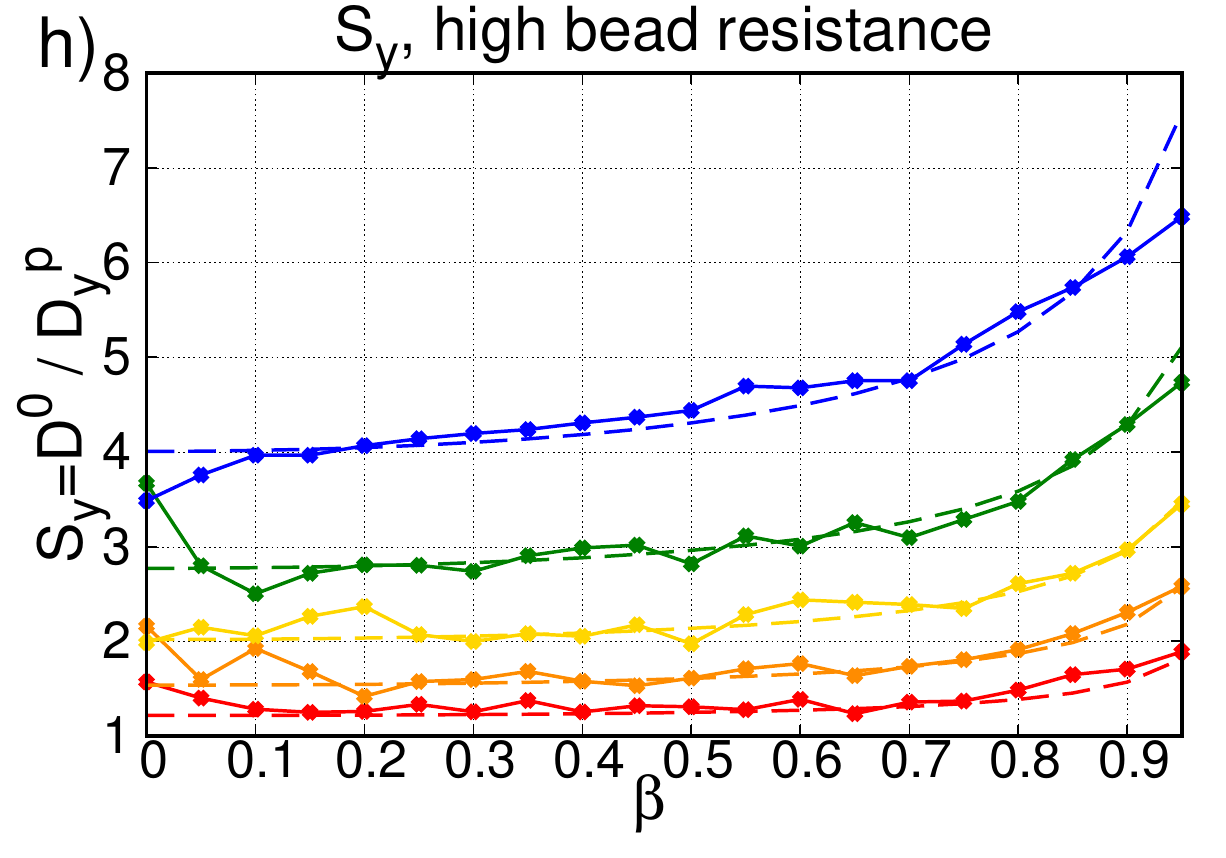}} &
		\resizebox{0.3\hsize}{!}{\includegraphics{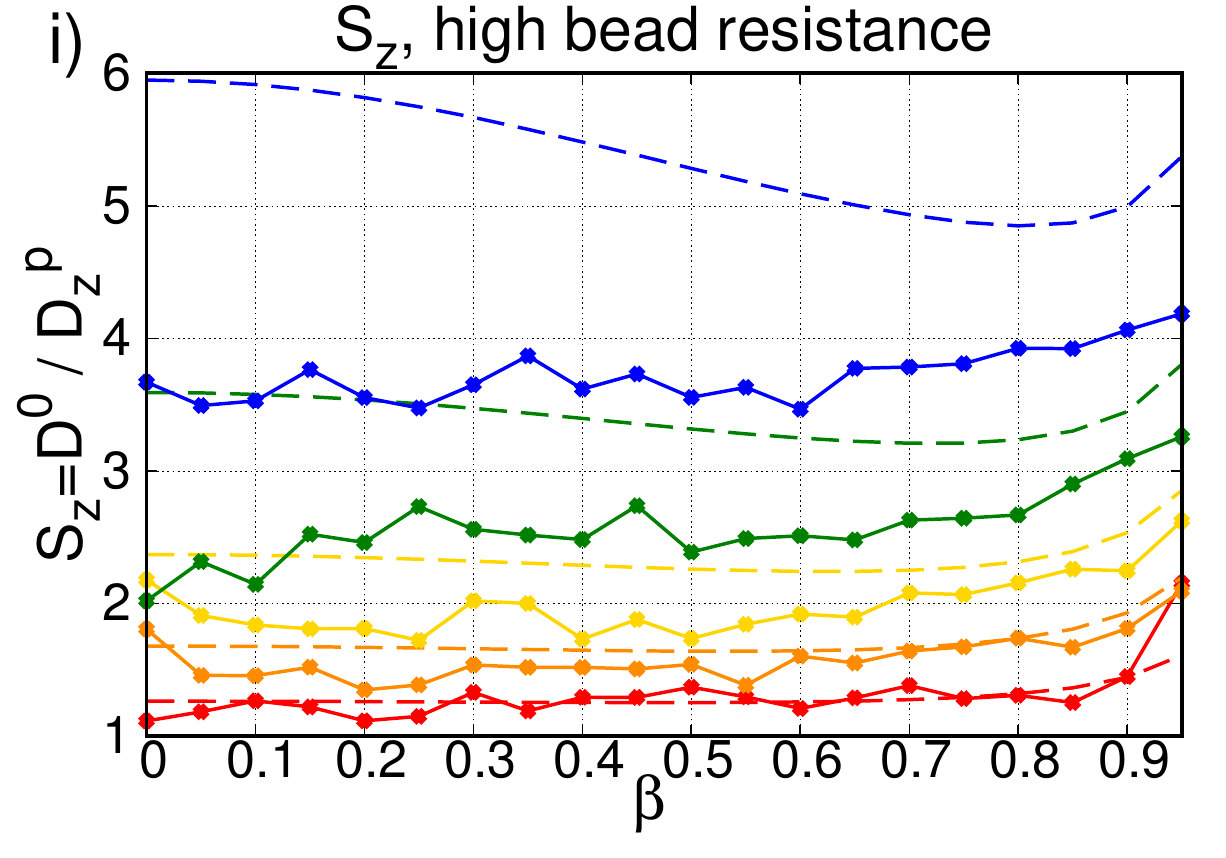}}\\
		\hline
	\end{tabular}
	\caption{Enhanced drag due to confinement within cylindrical pores. Red: $\lambda=0.1$. Orange: $\lambda=0.2$. Yellow: $\lambda=0.3$. Green: $\lambda=0.4$. Blue: $\lambda=0.5$. Top row (a,b,c): $\xi_{bead}=50M_{0}t^{-1}$. Middle row (d,e,f): $\xi_{bead}=200M_{0}t^{-1}$. Bottom row (g,h,i): $\xi_{bead}=1000M_{0}t^{-1}$. Left column (a,d,g): enhanced drag in $x^{p}$ direction. Middle column (b,e,h): enhanced drag in $y^{p}$ direction. Right column (c,f,i): enhanced drag in $z^{p}$ direction.}
	\label{figure:enhanced_drag}
\end{figure*}

%%%%%%%%%%%%%%%%%%%%%%%%%%%%%%%%%%%%%%%%%%%%%%%%%%%%%%%%%%%%%%%%
\subsection{Ellipsoids: lab-frame diffusion}
\label{sec:rsltslabdt}
The lab-frame translational and rotational diffusivities were\
determined via simulation for filled ellipsoidal raspberries of\
constant volume. After diffusivities were converted to translational and rotational\
resistances, the translational and rotational effective radii of the ellipsoids were calculated\
from the trendlines of $\xi^{t}_{col}$ versus $R_{rasp}$ and $\xi^{r}_{col}$ versus  $R_{rasp}^{3}$.\
The lab-frame geometric factors $A(\phi)$ and $B(\phi)$ of the colloid\
particles were determined using equations~\ref{eq:perrin_a} and~\ref{eq:perrin_b},\
respectively.

\begin{figure}
	\centering
	\resizebox{0.47\hsize}{!}{\includegraphics{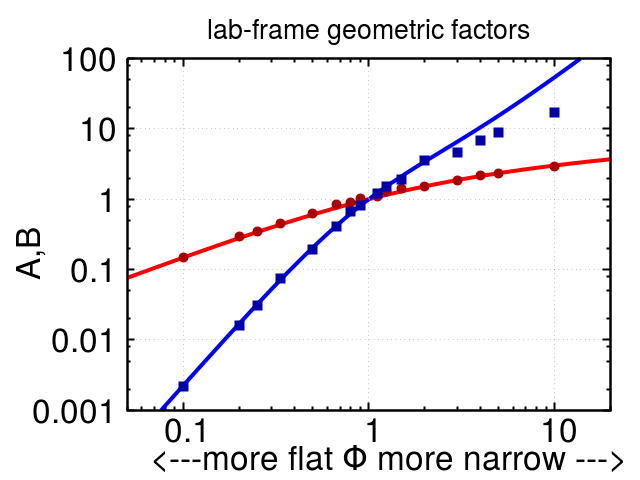}}
	\caption{Lab-frame transport geometric factors. The red line and blue line are the geometric factor values calculated by Perrin and represent A and B, respectively. Red circles and blue circles are the lab-frame geometric factors for the sixteen simulated ellipsoidal raspberries and represent A and B, respectively.}
	\label{figure:labframe_ellipsoids}
\end{figure}

Perrin~\cite{perrin1934,perrin1936} calculated the relationship between the translational and rotational effective radius $R_{eff}$, the aspect ratio $\phi$,\
and the singular semi-axis length $a$. The formulas can be found in the SI.\
The analytical geometric resistance factors\
were calculated by Perrin via averaging the friction over the entire body of the ellipsoid.\
Figure~\ref{figure:labframe_ellipsoids} shows these curves with respect to $\phi$. In prolate ellipsoids, $a$ is greater than\
$R_{eff}$, hence, $A(\phi)$ and $B(\phi)$ are both greater than 1.0. In oblate ellipsoids,\
$a$ is less than $R_{eff}$, hence, $A(\phi)$ and $B(\phi)$ are less than 1.0.

The lab-frame geometric factors $A(\phi)$ and $B(\phi)$ of the sixteen\
ellipsoids plus the curves for Perrin factors are presented in Figure~\ref{figure:labframe_ellipsoids}. Each of the sixteen\
ellipsoids are represented by one red circle and one blue square\
located at the same $\phi$ position on the plot. The translational\
factor $A(\phi)$ of these colloid particles are in excellent agreement\
with the analytical result.\
The rotational factor $B(\phi)$ is also in excellent agreement with\
Perrin's result for oblate ellipsoids and prolate ellipsoids for\
$\phi<3$. For prolate ellipsoids with $\phi>3$ the rotational\
diffusivity values are lower than Perrin's calculations.

%%%%%%%%%%%%%%%%%%%%%%%%%%%%%%%%%%%%%%%%%%%%%%%%%%%%%%%%%%%%%%%%%%%%
\subsection{Ellipsoids: body-frame diffusion}
\label{sec:rsltsbodyframe}
The body-frame translational and rotational diffusivities were\
determined for filled ellipsoidal raspberries of\
constant volume. The translational diffusivity of the ellipsoids in the body-frame\
followed the finite size scaling of $D_{s}$ versus $L^{-1}$.\
After conversion of the diffusivity values to colloid resistances, the $R_{eff,x^{b}_{i},t}$ for translation along the body axes\
and $R_{eff,x^{b}_{i},r}$ for rotation about the body axes were determined via the\
trendlines of $\xi^{t}_{col}$ versus $R_{rasp}$ and $\xi^{r}_{col}$ versus  $R_{rasp}^{3}$.

Happel and Brenner determined the geometric factors that define the\
relationship between an ellipsoid's actual body size $c$, aspect ratio\
$\phi$, and translational effective radius $R_{eff,x^{b}_{i},t}$ along\
the body-axes of an ellpsoid of revolution\
~\cite{happelbrenner1983}. Perrin determined the\
geometric factors that determine the rotational effective radius\
along different body-axes. The formulas can be found in the SI.

\begin{figure}
	\begin{tabular}{c c}
		\resizebox{0.47\hsize}{!}{ \includegraphics{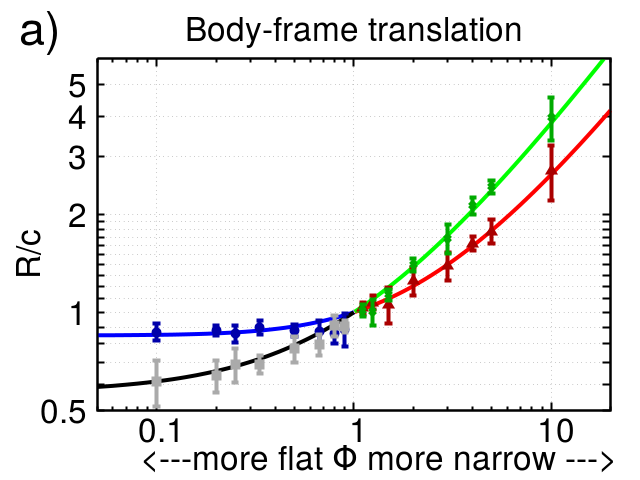}} &
		\resizebox{0.47\hsize}{!}{\includegraphics{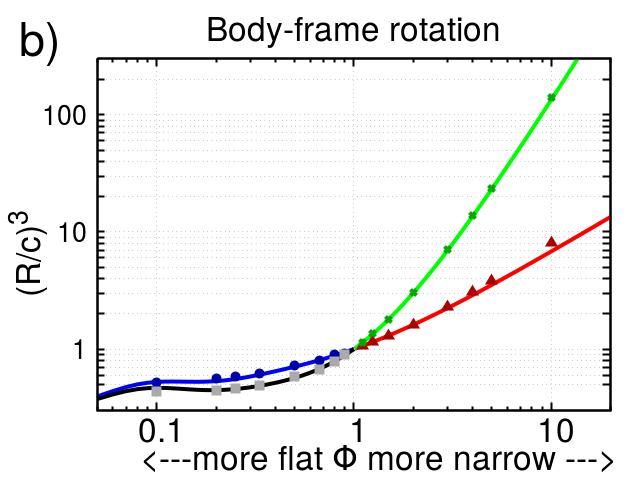}} \\
	\end{tabular}
	\caption{Body-frame transport geometric factors. a) Translation geometric factors. Red line, green line, blue line, and green line are the geometric factor values calculated by Happel and Brenner and represent $E$, $F$, $G$, and $H$, respectively. Red triangles, green x's, blue circles, and gray squares are the body-frame translational geometric factors for the simulated ellipsoidal raspberries and represent $E$, $F$, $G$, and $H$, respectively. b) Rotation geometric factors. Red triangles, green x's, blue circles, and gray squares are the body-frame rotational geometric factors for the simulated ellipsoidal raspberries and represent $I$, $J$, $K$, and $L$, respectively.}
	\label{figure:bftransport}
\end{figure}

Figure~\ref{figure:bftransport}.a gives the body-frame translational geometric\
factors $E$, $F$, $G$, $H$ of the sixteen ellipsoidal colloid particles\
plus the curves for the Happel-Brenner factors.\
The translational effective radii of prolate ellipsoids are smaller\
compared to the short body length $c$, hence, $E$ and $F$ are greater than 1.0.
The translational effective radii of oblate ellipsoids are larger\
compared to the short body length $c$, hence, $G$ and $H$ are less than 1.0. All\
ellipsoids agree very well with the Happel-Brenner factors.\
These data support the raspberry model for future applications of\
protein transport where anisotropy may help to explain open questions in protein\
separation~\cite{ku2004}.

The projection method to calculate body-frame angular displacement was tested for spheres to validate\
this method and can be found in the SI.\
Figure~\ref{figure:bftransport}.b gives the body-frame rotational geometric factors $I$, $J$, $K$, $L$ determined by simulation.\
The rotational effective radii of prolate ellipsoids are smaller\
compared to the short body length $c$, hence, $I$ and $J$ are greater than 1.0.\
The ratio of the rotational effective radii $\frac{R_{eff,a,r}}{R_{eff,bc,r}}=\left(\frac{I}{J}\right)^{1/3}$\
monotonically decreases with the aspect ratio of prolate ellipsoids.\
All ellipsoids agree very well with the Perrin factors~\cite{perrin1934,perrin1936}.

At long times, anisotropic transport reverts to isotropic transport\
via the diffusive rotation of an ellipsoid~\cite{han2006}. The time\
scale at which this occurs is $\tau_{\Theta}=\frac{1}{2D^{r}}=2.32$~ns-$1.16$~$\mu$s\
for the ellipsoidal raspberries in this study. In principle this\
crossover could be calculated by measuring many displacement\
trajectories by using one body-frame at a time to construct a full\
trajectory. Since these simulations are 2.541~$\mu$s in length,\
diffusion statistics using mean squared displacemens at lag times\
beyond a few nanoseconds are very poor. These measurements are\
therefore beyond the scope of this study. Interesting dissipative\
coupling between translational and rotational motion have also been\
observed for ellipsoids~\cite{han2006}, however, they manifest on the\
same $\tau_{\Theta}$ time scale and are beyond our reach.

\section{Conclusion}
\label{sec:conc}
The raspberry model in LB was investigated for protein-sized colloid\
particles. As bead resistance was increased, the overall colloid\
resistance to motion plateaued and attained a limiting value. The\
limiting colloid resistance to translation and rotation were higher\
than the predictions of the Stokes-Einstein and Stokes-Einstein-Debye\
relationships. The raspberries with high bead resistance were\
the least penetrable to LB as evidenced by significant enhanced drag under\
confinement in cylindrical pores. The enhanced drag was correct at high\
resistance values for coordinates more than a few LB grid spaces from the\
pore wall.

Since anisotropy has a pronounced effect on resistance to motion,\
ellipsoidal raspberries of aspect ratios between 0.1 and 10 were\
constructed and their transport was simulated. The Perrin and Happel-Brenner factors of these simulated\
colloids showed that the raspberry model reproduces the correct\
hydrodynamic resistances in the lab-frame and the body-frame.\

The raspberry model has been shown to be applicable to nonspherical colloid\
particles and appropriate for reproducing the hydrodynamics of protein-sized particles.\
It now allows us to go forward and use such a model in more complex\
environments where analytical calculations are not possible. Also more\
complex rigid shapes are now needed.

\section{acknowledgement}
Many helpful discussions with Ron Phillips, Pieter Stroeve, Joe\
Tringe, and Jonathan Higdon are gratefully acknowledged. We thank the UC Office of the\
President Labfee program (grant number 12-LR-237353) for financial\
support. We also thank Lawrence Livermore National Lab for allowing us\
access to their computer cluster.

\newpage
\section{SUPPLEMENTARY INFORMATION}

\section{Details for ellipsoidal raspberry construction}
\label{ellsurfconstruction}

See main text, Sec 2.1.

The type of surface dictates the complexity of the method required to calculate\
the bead-surface force. In the degenerate case of a sphere, a harmonic potential

\begin{equation}
\label{eq:harmonic_sphere}
U=k(r-R_{rasp})^2
\end{equation}

between the surface bead and the bead at the center of the sphere\
is sufficient to determine the force directed normal\
to the surface ~\cite{degraff2015rasp2layerconfined}.

\begin{figure}
	\centering
		\begin{tabular}{c c}
		\resizebox{0.28\hsize}{!}{ \includegraphics{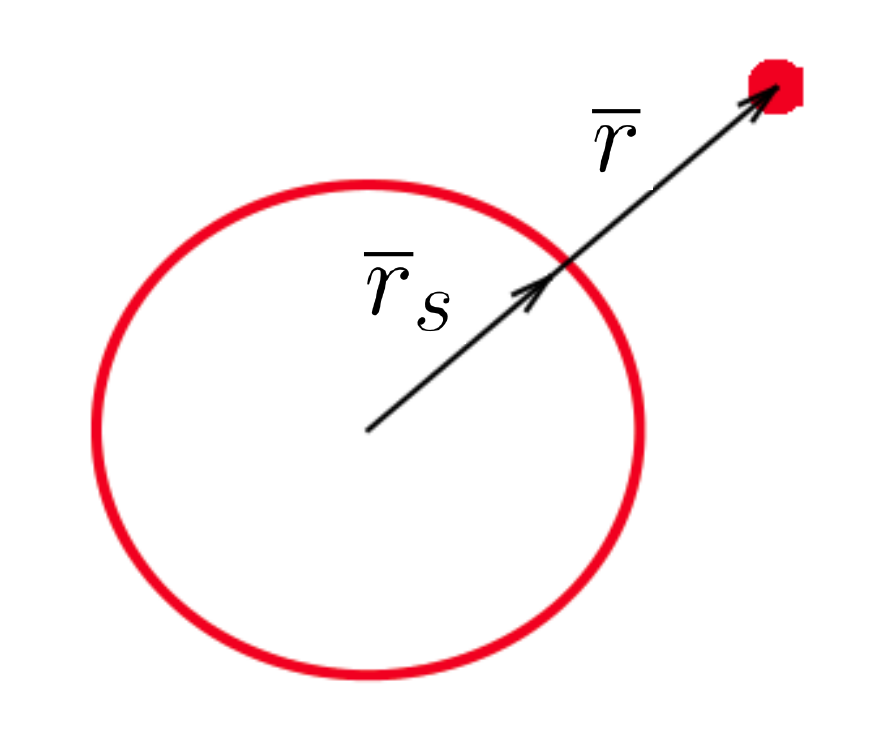} } &
		\resizebox{0.48\hsize}{!}{\includegraphics{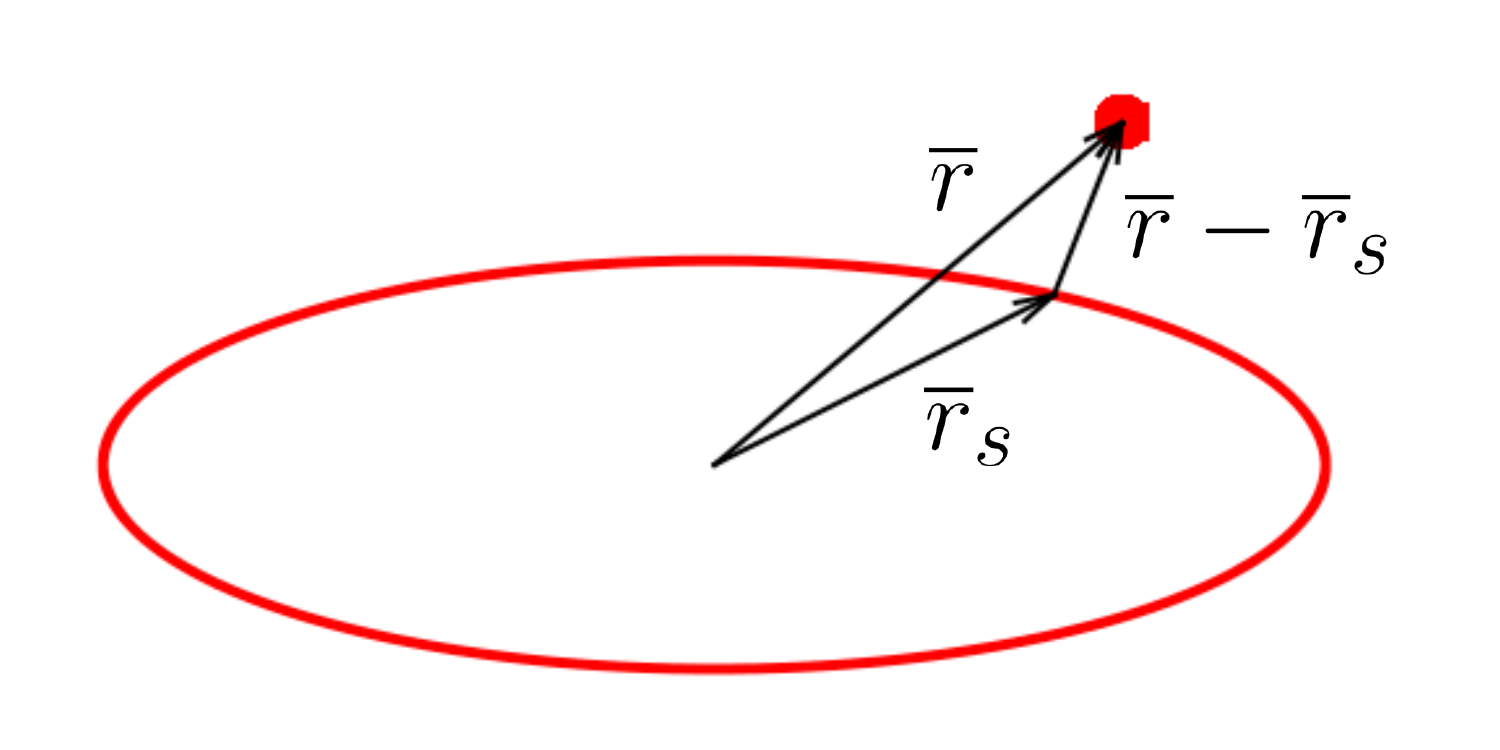} }\\
	\end{tabular}
	\caption{Cross section of sphere, left, and ellipsoid of revolution, right.\
	For an ellipsoid, the vector $\overline{r}-\overline{r}_{s}$ must\
	be determined in order to impose a surface restraining force on a bead at position\
	$\overline{r}$ towards surface position $\overline{r}_{s}$.}
	\label{figure:ell_surf}
\end{figure}

In the case of an anisotropic body,\
such as an ellipsoid of revolution $(a,b=c)$, where $a$, $b$, $c$ represent the lengths of the\
ellipsoidal semi-axes, the surface coordinate $\overline{r}_{s}=[x_{s}, y_{s}, z_{s}]$\
and the normal direction $\overline{n}$\
depend on polar and azimuthal location, as shown in a 2-D representation in\
Figure~\ref{figure:ell_surf}. For some particle coordinate $\overline{r}$ the surface coordinate $\overline{r}_{s}$ is therefore unknown and must be\
calculated numerically, which we describe here.

We start with the analytical description of the surface:

\begin{equation}
\label{eq:ellsurfacecoords}
\frac{x_s^2}{a^2}+\frac{y_s^2}{b^2}+\frac{z_s^2}{c^2} = 1
\end{equation}

From Figure 2 in the main text,

\begin{equation}
\label{eq:nandr}
	\frac{\overline{n}}{\|\overline{n}\|}=\frac{\overline{r}-\overline{r}_{s}}{\|\overline{r}-\overline{r}_{s}\|}.
\end{equation}

The normal vector $\overline{n}$ is also defined by the surface:

\begin{equation}
\label{eq:surfacegradient}
	\overline{n} = \nabla S = \left (\frac{2x_s}{a^2}\right)\overline{e}_x+\left (\frac{2y_s}{b^2}\right)\overline{e}_y+\left (\frac{2z_s}{c^2}\right)\overline{e}_z
\end{equation}

A parameter $t$ is defined:

\begin{equation}
\label{eq:tdefine}
	t \equiv 2 \frac{\|\overline{r}-\overline{r}_{s}\|}{\|\overline{n}\|}
\end{equation}

and substituted into Equation \ref{eq:nandr},

\begin{equation}
\label{eq:trearrange}
	\frac{t}{2} \overline{n} = \overline{r}-\overline{r}_{s}
\end{equation}

The surface gradient components in Equation \ref{eq:surfacegradient} are\
substituted into Equation \ref{eq:trearrange}:

\begin{equation}
	\frac{t}{2} \left [ \left (\frac{2x_s}{a^2}\right)\overline{e}_x+ \left (\frac{2y_s}{b^2}\right)\overline{e}_y + \left (\frac{2z_s}{c^2}\right)\overline{e}_z \right ] = \left (x-x_s\right)\overline{e}_x + \left (y-y_s\right)\overline{e}_y + \left (z-z_s\right)\overline{e}_z
\end{equation}

where the coordinates of a bead are $\overline{r}=[x,y,z]$.\
The relationship between the coordinates of the bead and the coordinates\
to the closest surface position are now established:

\begin{equation}
\label{eq:xtoxs}
	x=x_s \left ( \frac{t}{a^2}+1\right), y=y_s \left ( \frac{t}{b^2}+1\right), z=z_s \left ( \frac{t}{c^2}+1\right)
\end{equation}

by rearranging and inserting Equation \ref{eq:xtoxs} into Equation\
\ref{eq:ellsurfacecoords}, the function $F(t)$ is determined:

\begin{equation}
\label{eq:bisection}
	F(t) = \left ( \frac{xa}{t+a^2}+1\right)^2+\left ( \frac{yb}{t+b^2}+1\right)^2+\left ( \frac{zc}{t+c^2}+1\right)^2 -1 =0
\end{equation}

which can be solved using a bisection method. The largest value of\
$t$ that gives a solution to $F(t)<\lvert tol \rvert$ where\
$tol = 10^{-6}$ will give $\overline{r}_{s}$ using\
Equation \ref{eq:xtoxs}. Equation \ref{eq:bisection} was solved for\
each surface bead at each integration step to determine the\
surface constraining force during raspberry construction.

Ellipsoidal colloids were built in a script that integrated surface beads using the\
Leapfrog algorithm with a damping coefficient of $\eta = 1000$,\
Lennard-Jones size $\sigma = 1.4$, Lennard-Jones energy $\epsilon = 0.1$,\
time step $\Delta t = 0.001$. Typically $10^{7}$ integration steps were\
necessary to create a raspberry with evenly spaced surface beads. The\
surface density for all ellipsoids was one bead per nm$^{2}$ of surface area.
%%%%%%%%%%%%%%%%%%%%%%%%%%%%%%%%%%%%%%%%%%%%%%%%%%%%%%%%%%%%%%%%%%%%%%%%%
\section{Spheres: Radius}
\label{sec:radiusfit}

See main text, Sec. 3.2.

\begin{table*}
	\centering
	\begin{tabular}{lll}
		\hline\noalign{\smallskip}
		configuration                & $\xi^{t}_{col}$ vs. $R_{rasp}$                     & $\xi^{r}_{col}$ vs. $R_{rasp}^{3}$  \\
		\noalign{\smallskip}\hline\noalign{\smallskip}
		\scriptsize	theoretical                  &\scriptsize	 $\xi^{t}_{col}=6\pi\eta R_{rasp}=1.679\times10^{-2}R_{rasp}$    &\scriptsize	 $\xi^{r}_{col}=8\pi\eta R_{rasp}^{3}=2.224\times10^{-2}R_{rasp}^{3}$ \\
		\noalign{\smallskip}
		\scriptsize	low $\xi_{bead}$   &\scriptsize	 $\xi^{t}_{col}=2.047\times10^{-2}R_{rasp} - 1.424\times10^{-11}$ &\scriptsize	 $\xi^{r}_{col}=3.133\times10^{-2}R_{rasp}^{3} -9.539\times10^{-29}$ \\
		\noalign{\smallskip}
		\scriptsize	high $\xi_{bead}$ &\scriptsize	 $\xi^{t}_{col}=2.046\times10^{-2}R_{rasp} - 1.716\times10^{-12}$ &\scriptsize	 $\xi^{r}_{col}=4.173\times10^{-2}R_{rasp}^{3} + 2.640\times10^{-28}$ \\
		\noalign{\smallskip}\hline
	\end{tabular}
		\caption{Fits for colloid resistance to translational and rotational lab-frame transport. Units of slopes are kg~s$^{-1}$m$^{-1}$. Units of intercepts are kg~s$^{-1}$.}

	\label{tab:radiusfits}
\end{table*}

%%%%%%%%%%%%%%%%%%%%%%%%%%%%%%%%%%%%%%%%%%%%%%%%%%%%%%%%%%%%%%%%%%%%%%%%%
\section{Perrin factors for lab-frame transport of ellipsoids}
\label{sec:perrinfactors}

See main text, Sec. 2.4, 2.5, 3.4.

For an ellipsoid with semi-axis lengths $(a,b=c)$ and aspect ratio\
$\phi=a/b$, these factors are \cite{perrin1934,perrin1936}

\begin{equation}
\label{eq:a_phi}
	A(\phi) =\frac{a}{R_{eff,t}} = \frac{1}{\sqrt{|1-\phi^{-2}|}}\begin{cases}  \arctan\sqrt{\phi^{-2}-1} & \text{for oblate:  }\phi<1\\
    \ln\left ( \phi+\sqrt{\phi^{2}-1} \right ) & \text{for prolate: }\phi>1 \end{cases}
\end{equation}

\begin{equation}
\label{eq:b_phi}
	B(\phi) = \left ( \frac{a}{R_{eff,r}} \right )^3 =\frac{1+3\phi^{-2}A(\phi)}{2\phi^{-2}(1+\phi^{-2})}.
\end{equation}
%%%%%%%%%%%%%%%%%%%%%%%%%%%%%%%%%%%%%%%%%%%%%%%%%%%%%%%%%%%%%%%%%%%%%%%%%
\section{Happel-Brenner factors for body-frame translation of ellipsoids}
\label{sec:happelbrennerfactors}

See main text, Sec. 2.6, 3.5.

Note that the following equations are with respect to the degenerate\
axes $b$ and $c$, whereas the lab-frame equations for $A(\phi)$ and $B(\phi)$\
are written with respect to $a$. We have reported the equations as\
Happel and Brenner originally presented them \cite{happelbrenner1983}.

For translation along the singular axis of a prolate ellipsoid,

\begin{equation}
	\label{eq:happelbrenner_e}
	E(\phi) = \frac{R_{eff,x^{b},t}}{c} = \frac{8}{3}~ \frac{1}{ -\frac{2\phi}{\phi^{2}-1} + \frac{2\phi^{2}-1}{(\phi^{2}-1)^{3/2}} \ln \left ( \frac{\phi+\sqrt{\phi^{2}-1}}{\phi-\sqrt{\phi^{2}-1}} \right )}.
\end{equation}

For translation along the degenerate axes of a prolate ellipsoid,

\begin{equation}
	\label{eq:happelbrenner_f}
	F(\phi)=\frac{R_{eff,y^{b}z^{b},t}}{c}=\frac{8}{3}~ \frac{1}{ \frac{\phi}{\phi^{2}-1} + \frac{2\phi^{2}-3}{(\phi^{2}-1)^{3/2}} \ln \left ( \phi+\sqrt{\phi^{2}-1} \right )}.
\end{equation}

For translation along the singular axis of of an oblate ellipsoid,

\begin{equation}
	\label{eq:happelbrenner_g}
	G(\phi)=\frac{R_{eff,x^{b},t}}{c}=\frac{8}{3}~ \frac{1}{ \frac{2\phi}{1-\phi^{2}} + \frac{2-4\phi^{2}}{(1-\phi^{2})^{3/2}} \tan^{-1} \left ( \frac{\sqrt{1-\phi^{2}}}{\phi} \right )}.
\end{equation}

For translation along the degenerate axis of an oblate ellipsoid,

\begin{equation}
	\label{eq:happelbrenner_h}
	H(\phi)=\frac{R_{eff,y^{b}z^{b},t}}{c}=\frac{8}{3}~ \frac{1}{ -\frac{\phi}{1-\phi^{2}} + \frac{2\phi^{2}-3}{(1-\phi^{2})^{3/2}} \sin^{-1} \left ( \sqrt{1-\phi^{2}} \right )}.
\end{equation}
%%%%%%%%%%%%%%%%%%%%%%%%%%%%%%%%%%%%%%%%%%%%%%%%%%%%%%%%%%%%%%%%%%%%%%%%%

\section{Body-frame translational displacement}
\label{bftransdisplacement}

See main text, Sec. 2.6.

For an inverse quaternion $\overline{q}^{-1}(t) = (w,-x,-y,-z)$ at\
time $t$ the rotation can be written as a vector-matrix multiplication:

\begin{equation}
\label{eq:bodydisp}
	\delta \overline{X}^{b}_n = \overline{\overline{R}}^{-1} \delta \overline{X}_{n}
\end{equation}

where $\overline{\overline{R}}$ represents the rotation matrix. The\
notation is written as inverse because the inverse\
quaternion transformation is required here. The matrix is

\begin{equation}
\label{eq:quatmatrix}
\overline{\overline{R}}^{-1}=
\begin{pmatrix}
	1 - 2y^{2} - 2z^{2}  &  2xy - 2zw    &  2xz - 2yw     \\[0.3em]
	2xy + 2zw            & 1 - 2x^{2} - 2z^{2} &  2yz - 2xw     \\[0.3em]
	2xz - 2yw            & 2yz + 2xw     &  1 - 2x^{2} - 2y^{2} \\
\end{pmatrix}.
\end{equation}
%%%%%%%%%%%%%%%%%%%%%%%%%%%%%%%%%%%%%%%%%%%%%%%%%%%%%%%%%%%%%%%%%%%%%%%%%
\section{Body-frame angular displacement}
\label{sec:bfangulardisplacement}

See main text, Sec. 2.7, 3.5.

\begin{figure}[h]

	\centering
	\begin{tabular}{m{48mm}  m{48mm}  m{48mm} }
     \resizebox{1.0\hsize}{!}{ \includegraphics{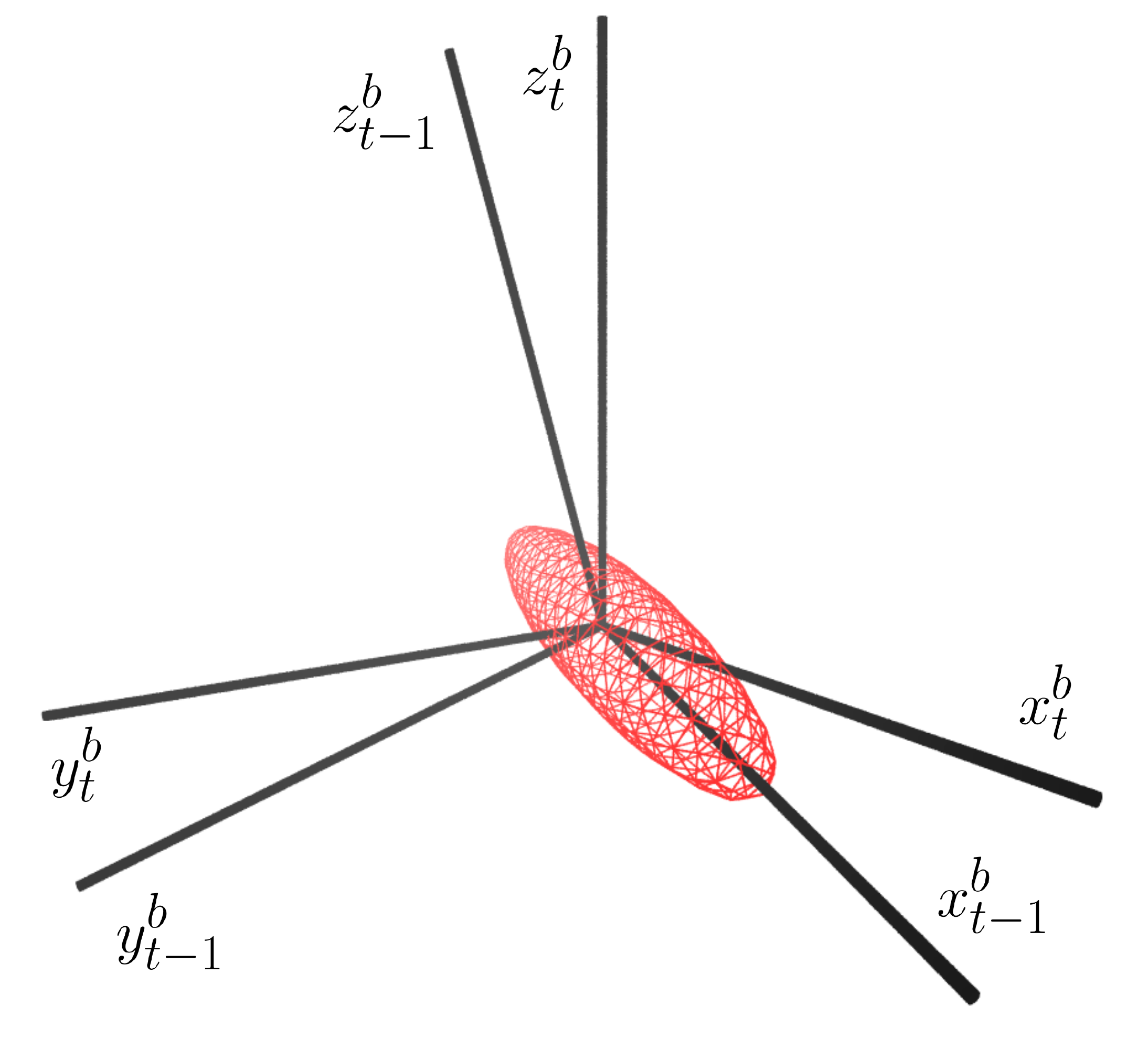}} &
     \resizebox{1.0\hsize}{!}{ \includegraphics{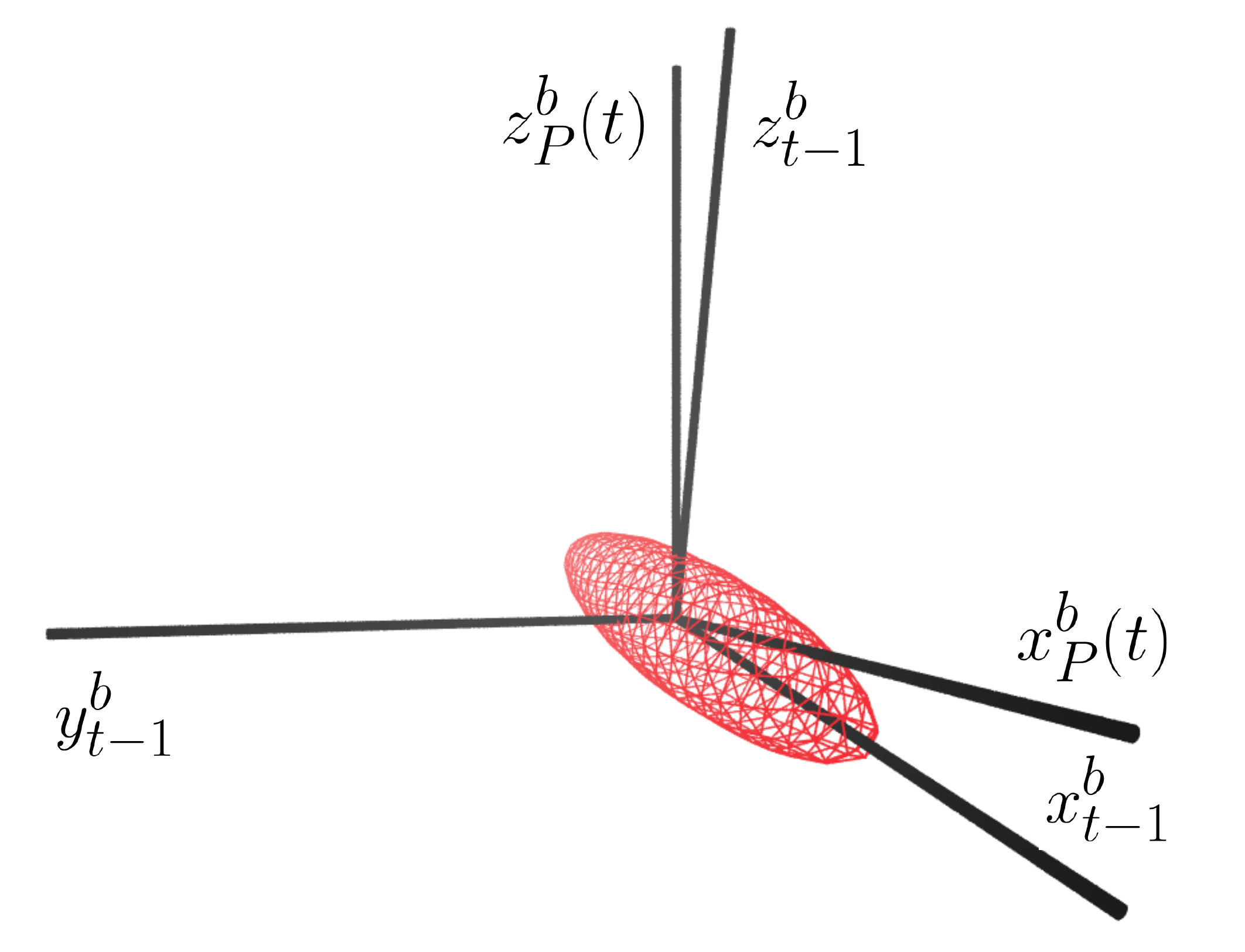} }&
     \resizebox{1.0\hsize}{!}{ \includegraphics{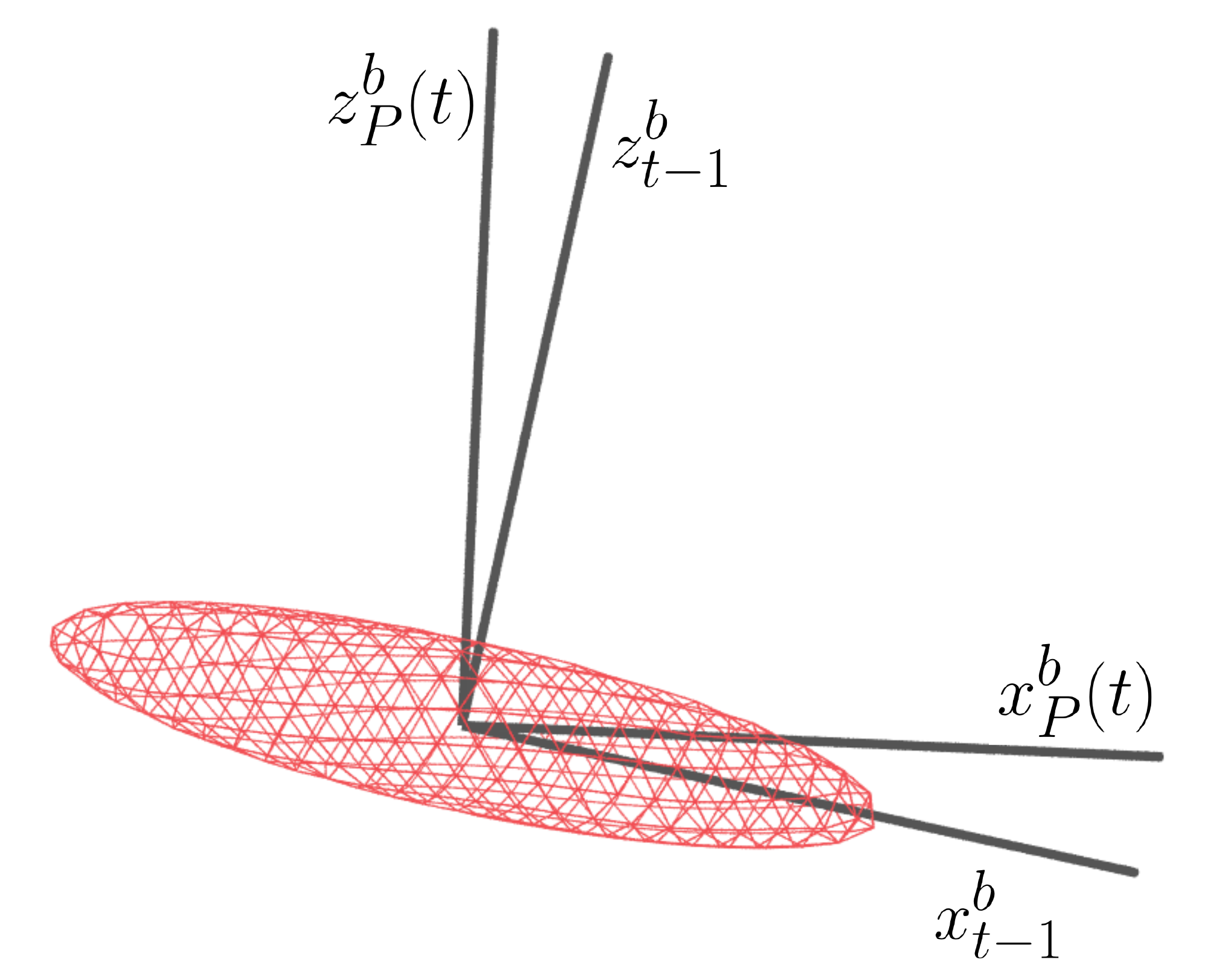} }\\
	\end{tabular}
	\caption{Projection method for measurement of body-frame rotation about $y^{b}$ between times $t$ and $t-1$. Left: Body frame axes for the $t$ and $t-1$ shapshot. Middle: Projections $x^{b}_{P}(t)$ and $z^{b}_{P}(t)$ of the $x^{b}_{t}$ and $z^{b}_{t}$ axes into the plane normal to $y^{b}_{t-1}$. Right: Same as middle image with perspective aligned with the $y^{b}_{t-1}$ axis, showing the angles between the $x^{b}_{P}(t)$ and $x^{b}_{t-1}$ which determines the angular displacement $\delta \Theta_{y^{b}}(t)$.}
	\label{figure:bfrot_axes}
\end{figure}

The body-frame rotations can be seen in Figure \ref{figure:bfrot_axes}.\
Quaternions were used to calculate the body-frame\
rotational motion.
The body-frame $x^{b}$ axis unit vector at time $t$ is

\begin{equation}
\label{eq:bodyframeunitvectorcalc}
\hat{i}^{b}_{t}= q_{t}\ \hat{i}\ q^{-1}_{t}=\overline{\overline{R}}\ \hat{i}
\end{equation}

and similarly calculated\
for the body-frame unit vectors $\hat{j}^{b}_{t} $ and $\hat{k}^{b}_{t} $. In\
order to observe a body-frame rotation about the $x^{b}$ axis between time $t-1$ and time $t$, the\
$\hat{k}^{b}_{t}$ vector was projected onto the $\hat{j}^{b}\hat{k}^{b}_{t-1}$\
plane:

\begin{equation}
	\hat{k}_{P}^{b}(t)=P_{\hat{j}^{b}\hat{k}^{b}_{t-1}}(\hat{k}_{t}^{b}) = \hat{k}_{t}^{b} - \hat{i}_{t-1}^{b}\left(\hat{k}_{t}^{b} \cdot \hat{i}_{t-1}^{b}\right).
\end{equation}

These projection vectors remain in the lab-frame, but in order to use\
arctangent to calculate the $\delta\Theta_{x^{b}}$, or the angle between\
$\hat{k}_{P}^{b}(t)$ and $\hat{k}^{b}_{t-1}$, transformation into the frame\
where $\hat{i}_{t-1}^{b}=\{1,0,0\}$,\ $\hat{j}_{t-1}^{b}=\{0,1,0\}$,\
$\hat{k}_{t-1}^{b}=\{0,0,1\}$ is required:

\begin{equation}
\hat{k}_{P}''(t) = \bar{\bar{A}}_{t-1}\ \hat{k}_{P}^{b}(t)
\end{equation}

The rotation matrix  $\bar{\bar{A}}_{t-1}$ is a standard rotation\
between two Cartesian coordinate frames:

\begin{equation}
\bar{\bar{A}}_{t-1}=
\begin{pmatrix}
	\hat{i}_{t-1}^{b} \cdot \hat{i} & \hat{i}_{t-1}^{b} \cdot \hat{j} & \hat{i}_{t-1}^{b} \cdot \hat{k}     \\[0.3em]
	\hat{j}_{t-1}^{b} \cdot \hat{i} & \hat{j}_{t-1}^{b} \cdot \hat{j} & \hat{j}_{t-1}^{b} \cdot \hat{k}     \\[0.3em]
	\hat{k}_{t-1}^{b} \cdot \hat{i} & \hat{k}_{t-1}^{b} \cdot \hat{j} & \hat{k}_{t-1}^{b} \cdot \hat{k}     \\
\end{pmatrix}
\end{equation}

where $\hat{i}=\{1,0,0\}$,\ $\hat{j}=\{0,1,0\}$,\
$\hat{k}=\{0,0,1\}$ are the unit vector lab-frame axes.
The angle displacement $\delta \Theta_{x^{b}}$ between time $t$ and\
$t-1$ is calculated via the arctangent function:\

\begin{equation}
\delta \Theta_{x^{b}}(t)=\arctan \left ( \frac{\hat{k}_{P2}''(t)}{\hat{k}_{P3}''(t)}      \right )
\end{equation}

where $P2$ and $P3$ refer to the 2nd and 3rd components of the\
rotated-projected vector, respectively. For displacements about the\
$y^{b}$ and $z^{b}$ axes, we calculate them using

\begin{equation}
\delta \Theta_{y^{b}}(t)=\arctan \left ( \frac{\hat{i}_{P3}''(t)}{\hat{i}_{P1}''(t)}      \right )
\end{equation}

\begin{equation}
\delta \Theta_{z^{b}}(t)=\arctan \left ( \frac{\hat{j}_{P1}''(t)}{\hat{j}_{P2}''(t)}      \right ).
\end{equation}
%%%%%%%%%%%%%%%%%%%%%%%%%%%%%%%%%%%%%%%%%%%%%%%%%%%%%%%%%%%%%%%%%%%%%%%%%
\section{Perrin factors for body-frame rotation of ellipsoids}
\label{poreframedetermination}

See main text, Sec. 2.7, 3.5.

Perrin calculated effective radii for rotation of a triaxial ellipsoid\
with semi-axis lengths $(a,b,c)$~\cite{perrin1934,perrin1936}. Note that the equations\
collapse for an ellipsoid of revolution $(a,b=c)$.\

For rotation about the singular axis of a prolate ellipsoid,

\begin{equation}
	\label{eq:dr_bf_i}
	I({\phi}) =  \left ( \frac{R_{eff,x^{b},r}}{c} \right )^{3}
\end{equation}

For rotation about the degenerate axes of a prolate ellipsoid,

\begin{equation}
	\label{eq:dr_bf_j}
	J({\phi}) =  \left ( \frac{R_{eff,y^{b}z^{b},r}}{c} \right )^{3}
\end{equation}

For rotation about the singular axis of an oblate ellipsoid,

\begin{equation}
	\label{eq:dr_bf_k}
	K({\phi}) =  \left ( \frac{R_{eff,x^{b},r}}{c} \right )^{3}
\end{equation}

For rotation about the degenerate axes of an oblate ellipsoid,

\begin{equation}
	\label{eq:dr_bf_l}
	L({\phi}) =  \left ( \frac{R_{eff,y^{b}z^{b},r}}{c} \right )^{3}
\end{equation}

The radii are calculated as

\begin{equation}
	\label{eq:r_rot_a}
	R_{eff,x^{b},r} =  \left ( \frac{2}{3Q} \right )^{1/3}
\end{equation}

\begin{equation}
	\label{eq:r_rot_a}
	R_{eff,x^{b},r} =  \left ( \frac{2}{3}~ \frac{a^{2} + c^{2}}{a^{2}P + c^{2}Q} \right )^{1/3}
\end{equation}

\begin{equation}
	\label{eq:P}
	P = \int^{\infty}_{0} \frac{ds}{(c^{2} +s)(a^{2} + s)^{3/2}}
\end{equation}

\begin{equation}
	\label{eq:Q}
	Q = \int^{\infty}_{0} \frac{ds}{(c^{2} +s)^{2}\sqrt{(a^{2} + s)}}
\end{equation}

%%%%%%%%%%%%%%%%%%%%%%%%%%%%%%%%%%%%%%%%%%%%%%%%%%%%%%%%%%%%%%%%%%%%%%%%%
\section{Pore-frame coordinate determination}
\label{poreframedetermination}

See main text, Sec. 2.8, 3.3.

The pore-frame coordinates of the raspberry are determined by\
translating the lab-frame origin to the center of the pore

\begin{equation}
	x^{c} = (x-0.5L_{xy})
\end{equation}
\begin{equation}
	y^{c} = (y-0.5L_{xy})
\end{equation}
\begin{equation}
	\theta = \arctan \left ( \frac{y_{c}}{x_{c}} \right ) .
\end{equation}

Over one interval,

\begin{equation}
\label{eq:xcenter}
\delta \overline{X}^{c}(t_{n})=
\begin{pmatrix}
	x^{c}(t_{n})-x^{c}(t_{n-1})   \\[0.3em]
	y^{c}(t_{n})-y^{c}(t_{n-1})  \\[0.3em]
	z(t_{n})-z(t_{n-1})  \\
\end{pmatrix} .
\end{equation}

The pore coordinates displacement are the result of a rotation

\begin{equation}
\delta \overline{X}^{p}(t_{n}) = \overline{\overline{P}}\ \delta \overline{X}^{c}(t_{n})
\end{equation}

via the following transformation matrix

\begin{equation}
\label{eq:porematrix}
\overline{\overline{P}}=
\begin{pmatrix}
	x_{1}\cos(\theta)  &  y_{1}\sin(\theta)    &  0     \\[0.3em]
	-x_{1}\sin(\theta)            & y_{1}\cos(\theta) &  0     \\[0.3em]
	0            & 0     &  1  \\
\end{pmatrix} .
\end{equation}

%%%%%%%%%%%%%%%%%%%%%%%%%%%%%%%%%%%%%%%%%%%%%%%%%%%%%%%%%%%%%%%%%%%%%%%%%
\section{Body-frame rotational diffusion of spheres}
\label{sec:bfrotspheres}

See main text, Sec. 3.5.

\begin{figure}[h!]
\centering
     \resizebox{0.5\hsize}{!}{ \includegraphics{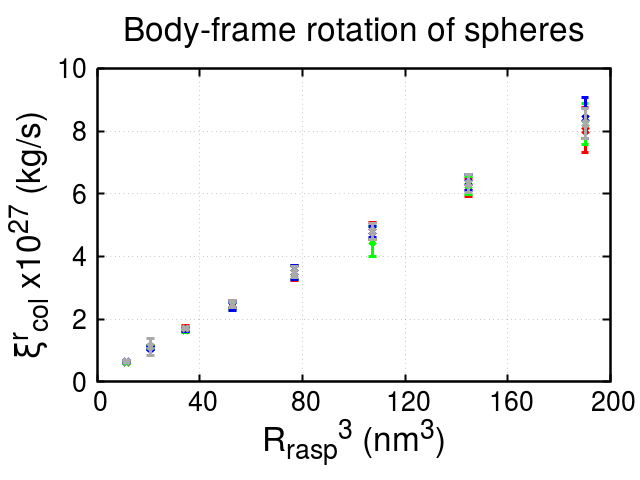} }
	\caption{Comparison of body-frame and lab-frame rotational resistance of filled spherical raspberries for different radii. Red, green, and blue x's represent colloid resistance to rotation about initial (arbitrary) body axes $(a=b=c)$ determined by the body-axis projection method. Gray x's represents resistance to rotation in the lab-frame.}
\label{figure:bfrotsphere}
\end{figure}

Figure \ref{figure:bfrotsphere} shows the rotational resistance of spherical raspberries\
versus size and validates the body-axis projection method for body-frame\
rotation. The mean-squared\
rotation about each of the $a$, $b$, and $c$ body axes are equivalent\
for all spheres.

%%%%%%%%%%%%%%%%%%%%%%%%%%%%%%%%%%%%%%%%%%%%%%%%%%%%%%%%%%%%%%%%%%%%%%%%%
\section{Enhanced drag for spheres with no outer layer}
\label{sec:bfrotspheres}
\begin{figure*}[h!]
	\begin{tabular}{|c c c|}
		\hline
		\resizebox{0.3\hsize}{!}{ \includegraphics{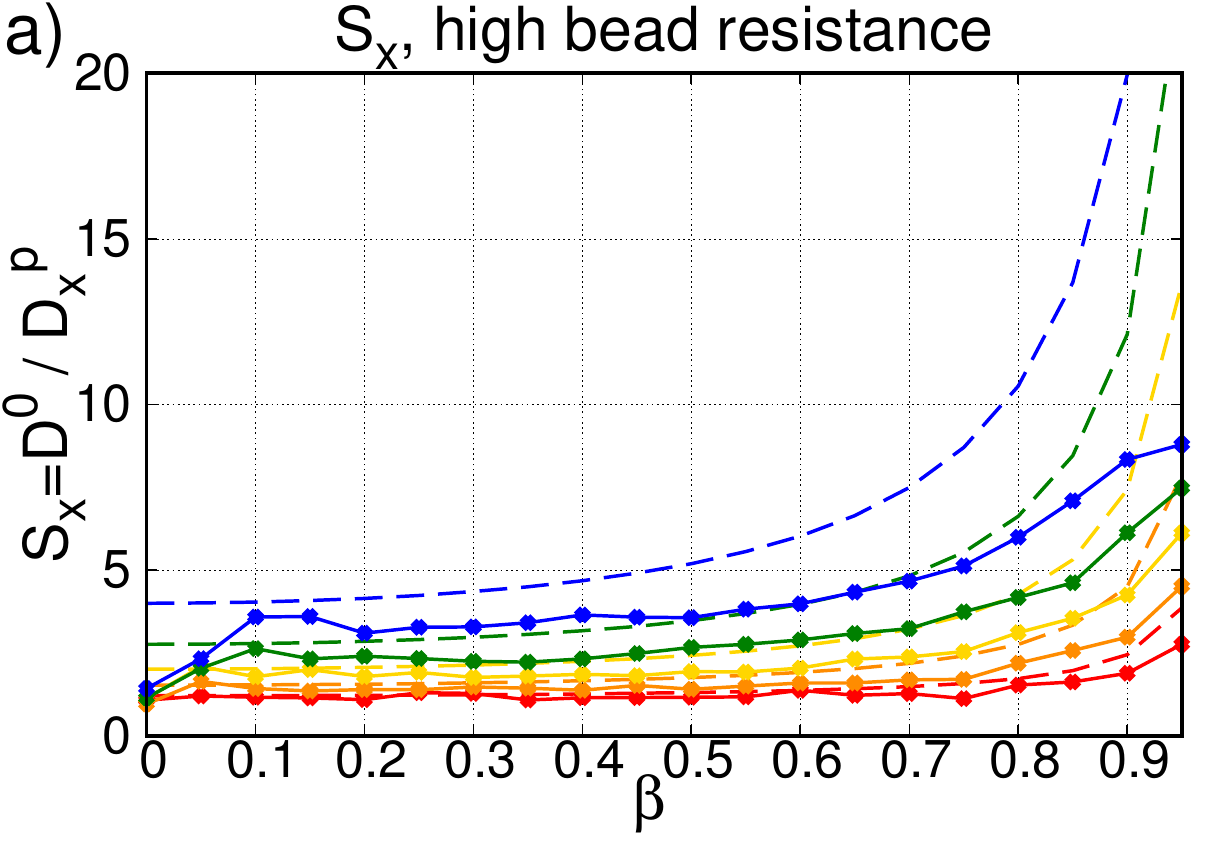}} &
		\resizebox{0.3\hsize}{!}{ \includegraphics{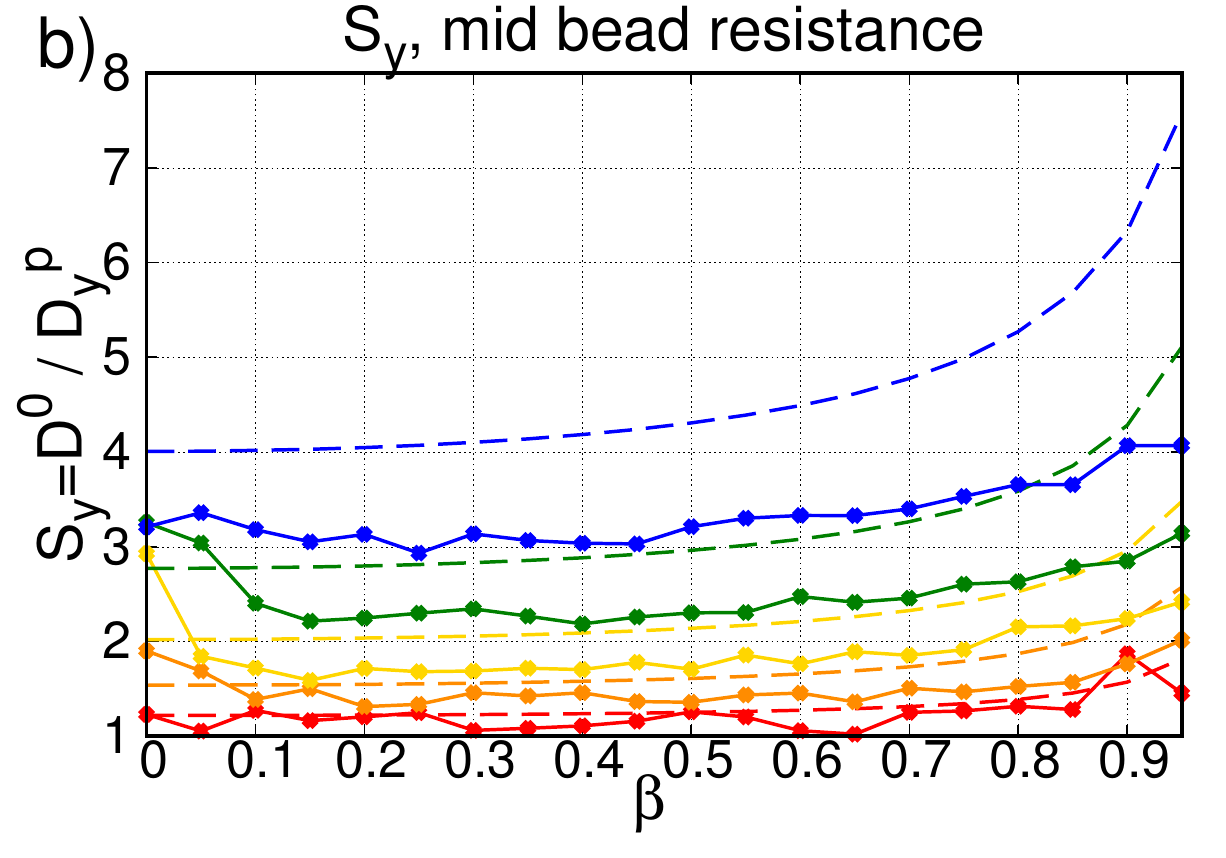}} &
		\resizebox{0.3\hsize}{!}{ \includegraphics{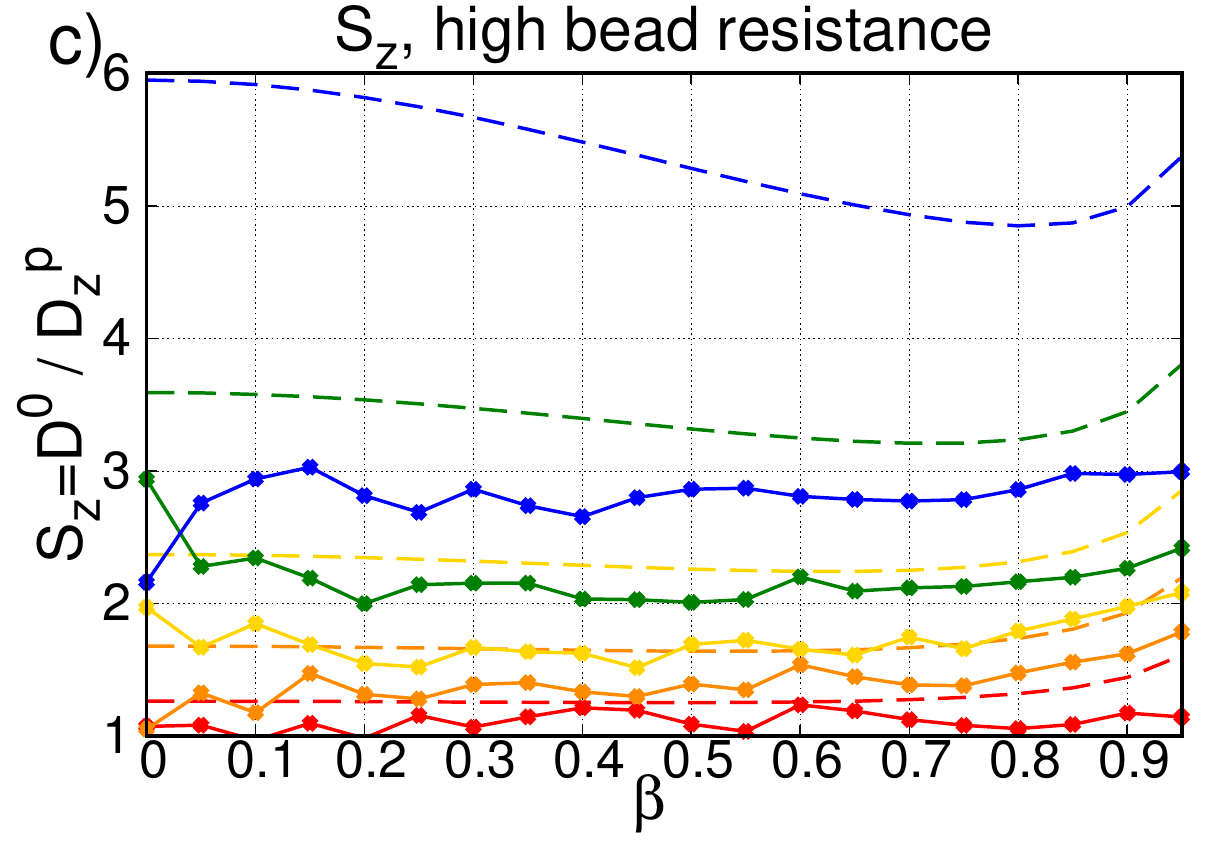}}\\
		\hline
	\end{tabular}
	\caption{Enhanced drag due to confinement within cylindrical pores, no outer layer, high bead resistance. Red: $\lambda=0.1$. Orange: $\lambda=0.2$. Yellow: $\lambda=0.3$. Green: $\lambda=0.4$. Blue: $\lambda=0.5$.}
	\label{figure:enhanced_drag}
\end{figure*}

%%%%%%%%%%%%%%%%%%%%%%%%%%%%%%%%%%%%%%%%%%%%%%%%%%%%%%%%%%%%%%%%%%%%%%%%%

\end{document}